\documentclass[aps,pra,reprint,superscriptaddress]{revtex4-2}

\usepackage{graphicx}
\usepackage{soul}
\usepackage{xcolor}
\usepackage{physics}
\usepackage{amsmath}
\usepackage{siunitx}
\usepackage{tabularx}
\usepackage{float}
\usepackage{booktabs}
\usepackage{array,multirow}
\usepackage{hyperref} 
\usepackage{orcidlink}
\usepackage{gensymb}
\usepackage[normalem]{ulem}

\hypersetup{
    colorlinks=true,
    linkcolor=blue,
    citecolor=blue,
    filecolor=blue,
    urlcolor=blue
}
\DeclareUnicodeCharacter{2212}{\textminus}

\newcolumntype{Y}{>{\raggedright\arraybackslash}X}

\newcommand{\redtransition}{{$^1$S$_0$}\,$\rightarrow$\,{$^3$P$_1$}}
\newcommand{\bluetransition}{{$^1$S$_0$}\,$\rightarrow$\,{$^1$P$_1$}}

\begin{document}

\preprint{APS/123-QED}

\title{Engineering Zeeman-manifold quintets using state-dependent light shifts in neutral atoms}

\author{Benedikt Heizenreder\,\orcidlink{0000-0001-5611-4144}}
\thanks{These authors contributed equally to this work.}
\affiliation{Van der Waals-Zeeman Institute, Institute of Physics, University of Amsterdam, Science Park 904, 1098 XH Amsterdam, The Netherlands}

\author{Bas Gerritsen\,\orcidlink{0009-0005-0496-4290}} 
\thanks{These authors contributed equally to this work.}
\affiliation{Institute for Theoretical Physics, Institute of Physics, University of Amsterdam, Science Park 904, 1098 XH Amsterdam, The Netherlands}
\affiliation{QuSoft, Science Park 123, 1098 XG Amsterdam, The Netherlands}

\author{Katya Fouka\,\orcidlink{0009-0002-8536-0955}}
\thanks{These authors contributed equally to this work.}
\affiliation{Institute for Theoretical Physics, Institute of Physics, University of Amsterdam, Science Park 904, 1098 XH Amsterdam, The Netherlands}
\affiliation{QuSoft, Science Park 123, 1098 XG Amsterdam, The Netherlands}

\author{Robert J. C. Spreeuw\,\orcidlink{0000-0002-2631-5698}} 
\affiliation{Van der Waals-Zeeman Institute, Institute of Physics, University of Amsterdam, Science Park 904, 1098 XH Amsterdam, The Netherlands}
\affiliation{QuSoft, Science Park 123, 1098 XG Amsterdam, The Netherlands}
\author{Florian Schreck\,\orcidlink{0000-0001-8225-8803}}
\affiliation{Van der Waals-Zeeman Institute, Institute of Physics, University of Amsterdam, Science Park 904, 1098 XH Amsterdam, The Netherlands}
\affiliation{QuSoft, Science Park 123, 1098 XG Amsterdam, The Netherlands}

\author{Arghavan Safavi Naini\,\orcidlink{0000-0002-5083-5423}} 
\affiliation{Institute for Theoretical Physics, Institute of Physics, University of Amsterdam, Science Park 904, 1098 XH Amsterdam, The Netherlands}
\affiliation{QuSoft, Science Park 123, 1098 XG Amsterdam, The Netherlands}

\author{Alexander Urech\,\orcidlink{0000-0003-1663-4705}}
\email{Quintet3P2@strontiumBEC.com} 
\affiliation{Van der Waals-Zeeman Institute, Institute of Physics, University of Amsterdam, Science Park 904, 1098 XH Amsterdam, The Netherlands}
\affiliation{QuSoft, Science Park 123, 1098 XG Amsterdam, The Netherlands}
\date{\today}

\begin{abstract}

We present a general method for engineering qudits through individually addressable transitions between Zeeman sublevels, achieved by combining a large linear Zeeman shift with a state-dependent light shift. This approach lifts the degeneracy between adjacent states while simultaneously tuning their energy splittings into the radio-frequency (RF) domain, enabling coherent manipulation within the Zeeman manifold using experimentally accessible drive frequencies. As a concrete realization, we investigate the implementation of an $SU(5)$ \emph{quintet} encoded in the Zeeman sublevels of the long-lived $^3\mathrm{P}_2$ state of neutral $\mathrm{^{88}Sr}$ atoms confined in far-detuned, $\sigma^{-}$-polarized optical tweezers. Using realistic experimental parameters, we numerically demonstrate full control of the \emph{quintet} manifold, including initialization into a specific $SU(5)$ basis state via a multi-photon transfer, coherent state- and site-selective single-qudit rotations driven by RF fields, and fast state-selective optical readout. Our simulations predict state-preparation fidelities of $\mathcal{F} \simeq 0.99$ within less than $1~$\si{\micro\second}, single-qudit gate fidelities of $F$ $\simeq 0.99$ with $\pi$-pulse durations of $\sim1.4~$\si{\micro\second}, and fast destructive imaging with durations below $10~$\si{\micro\second} under ideal conditions. These results establish a broadly applicable framework for high-fidelity control of Zeeman sublevel-encoded qudits and highlight the $^3\mathrm{P}_2$ manifold in strontium as a promising platform for scalable qudit-based quantum technologies.
\end{abstract}

\maketitle            

\section{Introduction}
\label{sec:Introduction}
Neutral alkaline-earth-metal atoms trapped in arrays of optical tweezers or optical lattices form a powerful platform for atomic clocks and sensors~\cite{cao_multi-qubit_2024, eckner_realizing_2023, young_half-minute-scale_2020}, quantum computing~\cite{reichardt_logical_2024, muniz_high-fidelity_2024, lis_mid-circuit_2023}, and quantum simulation of many-body physics~\cite{madjarov_high-fidelity_2020, tao_high-fidelity_2024, holman_trapping_2024}. Their success is largely driven by the coherent control of optical intercombination transitions, exploiting extremely long-lived excited states that enable coherence times on the order of minutes~\cite{Kim2025_WannierStarkClock}. At the same time, these advantages require highly coherent and long-term stable laser systems~\cite{kedar_frequency_2023,yu_excess_2023}. Moreover, the need for multiple fast rotations and sophisticated decoupling sequences in the optical domain remains a significant experimental challenge~\cite{Milner2025_SuperexchangeClock, Yan2025_ClockLaserQudit}.

Parallel to these developments, there is a growing interest in exploiting higher-dimensional Hilbert spaces~\cite{lindon_complete_2023, sonderhouse_thermodynamics_2020, Yamamoto_Quantum_and_Thermal_Phase, Nataf_Exact_Diagonalization}.
Coherent control over multiple internal atomic levels enables the realization of so-called \emph{qudits}. 
In recent years, there have been theoretical and experimental efforts to develop qudit systems for simulation of lattice gauge theories~\cite{Meth2025,Gaz2025,Popov2024, Illa2023,Gonzalez2022,Ciavarella2021,Calajo_2024,Gonzalez-Cuadra2025} and higher spin models such as the AKLT model ~\cite{Edmunds2025, Mogerle2025, Wang2023, Senko2015, Cohen2015, Cohen2014}. Beyond their role in quantum simulation, qudit systems are also anticipated to provide significant advantages for quantum-enhanced metrology and sensing~\cite{Ma_2026,Kaubruegger2023,Javaherian_2025, Ahmed2025,lin2026} as well as for quantum computing, where they may lead to more scalable fault-tolerant architectures~\cite{Campbell2014_EnhancedFaultTolerant_dLevel, Lim2023_Spin7_2Qudit} and reduced circuit complexities~\cite{Lanyon2009}. Finally, by allowing multiple logical qubits to be encoded within a single atom, qudits can decrease the underlying resource requirements~\cite{omanakuttan_qudit_2023,Gross2024_HardwareEfficient,Lim2025_ErrorCorrectable_Qudit,Mezzadri2024_FaultTolerantQudit}.

One can use different approaches to realize such qudits in neutral atoms. One strategy encodes information in nuclear spins of the electronic ground state, utilizing their exceptional isolation from the environment and long coherence times~\cite{nakamura_hybrid_2024, jia_architecture_2024, Coherent_Control_Over_High-Dimensiona_Ahmed, jenkins_ytterbium_2022}. Alternatively, metastable electronic sub-levels~\cite{omanakuttan_qudit_2023, Schuetz_3P2_Neon,ammenwerth_realization_2024,unnikrishnan_coherent_2024} or even a combination of nuclear spins and electronic levels provide attractive opportunities~\cite{Huie2025_threequbit}. Particularly appealing are qudits where computational coupling is performed in the radio-frequency (RF) domain~\cite{omanakuttan_quantum_2021, omanakuttan_qudit_2023, Schuetz_3P2_Neon}. The long RF wavelength enables phase-coherent control, precise and fast rotations, and the implementation of demanding dynamical decoupling sequences~\cite{Soonwon_Dynamical_Engineering, Souza_Robust_Dynamical, Choi_Robust_Dynamic_Hamiltonian}. These operations are relatively easy to implement compared to the optical domain. However, while high-fidelity control in the RF domain is well established, preparation and readout of the qudit must still be performed optically, and scaling such methods to systems with more than two levels remains a key challenge~\cite{Coherent_Control_Over_High-Dimensiona_Ahmed, Schuetz_3P2_Neon,ammenwerth_realization_2024}.  

\begin{figure}[]
\centering
\includegraphics[scale=1.0]{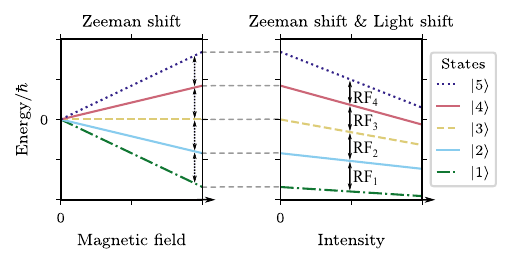}
\caption{
Linear Zeeman shifts, shown in the left Breit-Rabi level diagram, lift the degeneracy between the five targeted Zeeman sublevels and tune the energy differences between neighboring states, by the applied magnetic field, into a range that is easily accessible with standard RF technology ($\sim 100~\mathrm{MHz}$). 
The energy differences (or corresponding RF drive frequencies) between neighboring states, indicated by dashed black arrows, remain degenerate under the Zeeman shift alone. 
By combining the constant Zeeman shift with a state-dependent light shift, induced through the tweezer beams and linearly dependent on the intensity (illustrated in the right panel), the degeneracy is lifted, enabling individual RF drive frequencies for all four transitions (solid arrows). 
}
\label{fig:Breit_Rabi_combined}
\end{figure}

In this work, we study the realization of an $SU(5)$ \emph{quintet} encoded in neighboring Zeeman sublevels. To lift the degeneracy between these levels, we combine a large linear Zeeman shift, induced by an external magnetic field, with a state-dependent light shift generated by the $\sigma$-polarized optical tweezer beams (see Fig.~\ref{fig:Breit_Rabi_combined}). This approach engineers non-degenerate energy splittings between adjacent states, allowing individual RF drive frequencies within a range easily accessible to standard RF technology ($\sim 100~\mathrm{MHz}$). We specifically explore a potential experimental realization (Sec.~\ref{sec:Potential_experimental_realization}) that benefits from the long lifetime (tens of seconds) of the $^3$P$_2$ state in neutral $\mathrm{^{88}Sr}$ atoms confined in far-off resonantly detuned optical tweezers. This system offers near-perfect isolation from the main imaging and cooling transitions while remaining compatible with established RF spin-control techniques developed for alkali atoms. We numerically simulate fast state preparation (Sec.~\ref{sec:state_preparation}), the feasibility of high-fidelity quintet rotations (Sec.~\ref{sec:RF-drive}), state-selective readout, and rapid high-fidelity imaging under experimentally realistic parameters (Sec.~\ref{sec:SSreadout}). Finally, we discuss possible routes to further improve rotation fidelities, achieve a universal gate set, and increase circuit depth (Sec.~\ref{sec:Outlook}).

The presented study shows that the $^3$P$_2$ quintet in $\mathrm{^{88}Sr}$ offers a practical and scalable route to implement qudit-based quantum technologies, combining long coherence with fast state preparation and readout, precise rotation control, and rapid high-fidelity imaging.  

\begin{figure}[]
\centering
\includegraphics[width=8.3cm]{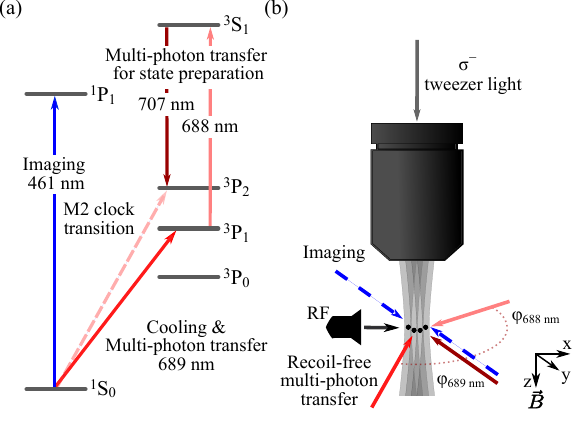}
\caption{(a) Level diagram of $\mathrm{^{88}Sr}$ showing the blue imaging and red cooling transitions~\cite{Pucher2025_88Sr_Reference}, together with the states involved in the multi-photon transfer used for initial-state preparation and readout of the $^3$P$_2$ Zeeman manifold \emph{quintet}.
(b) Proposed experimental setup, consisting of a high-resolution imaging system for preparing $\sigma^-$-polarized optical tweezers with single-atom occupancy at high magnetic field. An RF antenna generates the field required to couple neighboring Zeeman sublevels, which serve as computational basis states. Additional laser beams are indicated for imaging, cooling, and the multi-photon transfer, where $\varphi_{689\mathrm{nm}}$ and $\varphi_{688\mathrm{nm}}$ denote the in-plane angles between the coupling beams chosen to minimize the net momentum transfer, $\hbar |\mathbf{k}_{\mathrm{eff}}| \approx 0$.}
\label{fig:setup_qudit}
\end{figure}

\section{$^3$P$_2$ quintet}
\subsection{Proposed experimental realization}
\label{sec:Potential_experimental_realization}
The high natural abundance ~\cite{Haynes2016} and the simple and well-characterized level structure of $\mathrm{^{88}Sr}$ (see Fig.~\ref{fig:setup_qudit}(a)) make it an ideal candidate for realizing the $^3$P$_2$ quintet. Here, we propose an experimental setup consisting of a one-dimensional optical tweezer array generated by an acousto-optic deflector (AOD) in combination with a high-resolution imaging system (NA = 0.5) to allow for single-atom control~\cite{Shaw_Dark-State_Enhanced}. Employing a tweezer wavelength of $1064~\mathrm{nm}$ with $\sigma^-$ polarization yields a waist of approximately $1~$\si{\micro\meter} (see Fig.~\ref{fig:setup_qudit}(b)). Such far-off-resonant tweezers support atom lifetimes of several seconds for the $^3$P$_2$ metastable state even at millikelvin trap depths, due to the favorable polarizability of all $^3$P$_2$ metastable states (see App.~\ref{app:Scattering}). We note that while the availability of high-power lasers at this wavelength makes $1064~\mathrm{nm}$ a practical choice, any other wavelength in the range $800$–$1100~\mathrm{nm}$ could be suitable (see App.~\ref{app:Polarizability}). 

\begin{figure}[]
    \centering
    \includegraphics[width=8.3cm]{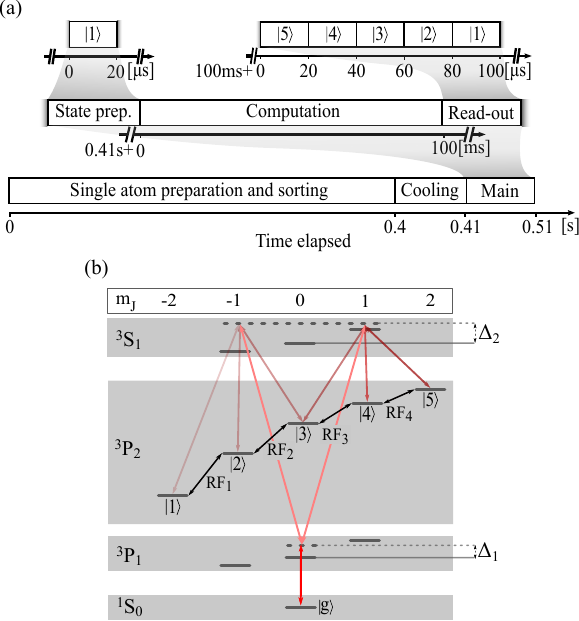}
    \caption{
    (a) Envisioned timing sequence following initial single-atom preparation and cooling. The full experimental cycle is divided into three stages: State preparation, computation, and final readout. 
    (b) State preparation and readout: atoms are transferred from the ground state $^1$S$_0$ to the $^3$P$_2$, $m_J=1$ state via a multi-photon process, thereby initializing the system in one of the computational basis states. As an example, the bare detunings $\Delta_1$, $\Delta_2$ are shown for this $\ket{g}$\,$\leftrightarrow$\,$\ket{1-5}$ transition. $\Delta_3$ is defined in Eq.~(\ref{eq:multi_con}) but not shown in the figure. After computation, the same multi-photon transfer selectively transfers population from the $^3$P$_2$ manifold back to the ground state, where each state is destructively imaged. For clarity, only $\pi$-polarized light for the $^1$S$_0$\,$\leftrightarrow$\,$^3$P$_1$ transition and the $^3$P$_1$, $m_J=0$ intermediate state are depicted (also energy levels are not to scale); however, all polarizations and possible transitions are included in the calculations. Note that the $^3$S$_1$, $m_J=0$ to $^3$P$_1$, $m_J=0$ transition is forbidden by angular-momentum selection rules. 
 A combination of Zeeman and tensor light shift lifts the degeneracy of the $^3$P$_2$ Zeeman sublevels, enabling individual addressing of transitions (black solid arrows) using an RF field.
}
\label{fig:pumping_schema}
\end{figure}

The choice of a 1D array controlled by an AOD enables site-specific, rapid control of the individual tweezer intensities and positions, which is useful for single-atom preparation and sorting~\cite{Urech_Narrow-line_imaging,Verstraten2025_control,Norcia2018_Sr_Tweezers_GSC}, as well as for single-site addressability and trap depth balancing discussed later in Sec.~\ref{sec:RF-drive_single_site}. The timing sequence for the full experimental cycle is shown in Fig.~\ref{fig:pumping_schema}(a). Moreover, the millikelvin-deep $\sigma^-$-polarized tweezers considered here provide near-magic trapping conditions for the cooling transition (\redtransition, $m_J=+1$). Since the tweezers create large trap frequencies, one can utilize sub-Doppler cooling techniques such as resolved sideband cooling or attractive Sisyphus cooling in both the axial and radial directions.  Thus, we envision that such deep, far off-resonant traps will enable preparation of atoms near the motional ground state, with mean occupations close to $\bar{n}_{x,y} \simeq 0.01$ in the radial directions and $\bar{n}_z \simeq 0.1$ axially for the $^1$S$_0$ ground state, as demonstrated in comparable systems~\cite{Norcia2018_Sr_Tweezers_GSC,young_half-minute-scale_2020}. Additionally, state-of-the-art techniques such as erasure cooling~\cite{Shaw_ErasureCooling} can ensure near-perfect ground state cooling for metastable state atoms, below our cooling simulation results presented here.

Throughout this work, we consider a static homogeneous magnetic field of $100~\mathrm{G}$, aligned with the tweezer propagation direction ($z$-axis in Fig.~\ref{fig:setup_qudit}(b)). This field defines the quantization axis and splits the $^3$P$_2$ manifold into five resolvable Zeeman sublevels with energy separations of approximately $210~\mathrm{MHz}$ between neighboring $m_J$ states. We consider magnetic field fluctuations to be negligible in this regime with proper field stabilization as previously demonstrated to the ppm level~\cite{ActivestabilizationofkGField,subPPmfieldstabilityat100G}, allowing for coherence times on the level of tens of miliseconds or more. We lift the degeneracy between the $^3$P$_2$ Zeeman transition frequencies with $\sigma^-$-polarized tweezer light (see Fig.~\ref{fig:pumping_schema}(b)). The associated tensor light shift differentiates transitions between neighboring $^3$P$_2$ $m_J$ states, enabling individual addressability via unique RF transition frequencies. These RF fields, which can be generated by an RF antenna oriented and polarized orthogonally to the tweezer axis (see Fig.~\ref{fig:setup_qudit}(b)), provide the coupling required for computation within the quintet, where the five $m_J$ states serve as the computational basis. The resulting unique transition frequencies enable high-fidelity Rabi oscillations between adjacent $m_J$ states (Sec.~\ref{sec:RF-drive}). The frequency shifts are sufficient to resolve individual transitions even at Rabi frequencies on the order of $2\pi \times 200~\mathrm{kHz}$ when using deep tweezers with tens of milliwatts of power per beam.
We envision the tweezer power to be stabilized (short-term) to the $0.02 \%$ level~\cite{toptica_1064nm}, making power fluctuations on the timescale of a single gate litigable. However, we do take into account slower shot-to-shot drifts or power inhomogeneities across a qudit array.

For initial state preparation and final state-selective readout (see Fig.~\ref{fig:pumping_schema}(b)), we simulate a multi-photon transition $^1$S$_0$\,$\leftrightarrow$\,$^3$P$_2$ proceeding via $^1$S$_0$\,$\rightarrow$\,$^3$P$_1$\,$\rightarrow$\,$^3$S$_1$\,$\rightarrow$\,$^3$P$_2$~\cite{ammenwerth_realization_2024,He_Coherent_three-photon}. This scheme avoids the weak direct coupling between $^1$S$_0$ and $^3$P$_2$, allowing for effective Rabi frequencies at the MHz level (see Sec.~\ref{sec:state_preparation} and Sec.~\ref{sec:SSreadout}). Compared to using a single ultra-stable laser, this approach also offers the advantage of efficiently coupling to all $^3$P$_2$ states with high Rabi frequencies~\cite{ammenwerth_realization_2024,Pucher_Fine-Structure}. 
Furthermore, by directing the coupling beams orthogonal to the tweezer axis and finely adjusting their relative angles to $\varphi_{689\mathrm{nm}} \approx -61\degree$ and $\varphi_{688\mathrm{nm}} \approx 61\degree$ (defined with respect to the $707\mathrm{nm}$ beam; see Fig.~\ref{fig:setup_qudit}(b)), the net momentum transfer $\hbar |\mathbf{k}_{\mathrm{eff}}|$ induced through the multi-photon transfer can be canceled~\cite{Panelli_Doppler-free_three-photon_2025,Barker_Three_photon_2016}. 
Under these conditions, and considering the near-magic trapping for the 
{$^1$S$_0$}\,$\rightarrow$\,{$^3$P$_2$\, $m_J=1$} transition, we assume that 
after initial state preparation, the atoms remain close to the motional ground 
state, with mean occupations of $\bar{n}_{x,y,z} = \{0.01, 0.01, 0.20\}$ in the 
initialized state (see App.~\ref{app: Minimizing motion}).
Finally, we simulate fast, destructive imaging on the $^1$S$_0\,\leftrightarrow\,^1$P$_1$ transition using chopped, counter-propagating beams orthogonal to the tweezer axis, as discussed in Sec.~\ref{sec:SSreadout} and App.~\ref{app:Fast imaging}.

\subsection{State preparation}
\label{sec:state_preparation}

\begin{figure}[]
    \centering
    \includegraphics[scale=1.0]{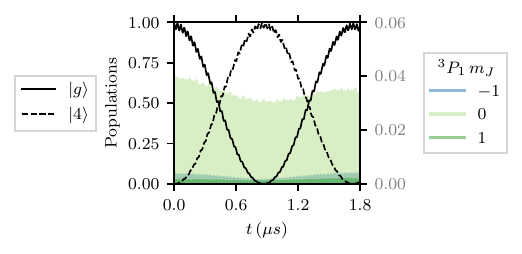}
    \caption{State preparation of $^3$P$_2, m_J=1$. Bold left axis and lines: Dynamics of the {$^1$S$_0$} ($\ket{g}$) $\leftrightarrow$\,{$^3$P$_2, m_J=1$} ($\ket{4}$) transition, including the Zeeman substructure, under the multi-photon resonance condition for {$^3$P$_2, m_J=1$} [Eq.\eqref{eq:multi_con}]. Faint right axis and colored lines:  magnified view, showing rapid oscillations of the $^3\mathrm{P}_1$ populations (not resolvable on this time scale). Parameters: 
    $\Omega_1/2\pi = $ 10.9 MHz, $\Omega_2/2\pi = $ 1380.6 MHz, $\Omega_3/2\pi$ = 219.9 MHz, $\Delta_1/2\pi$ = 10.8 MHz, $\Delta_2/2\pi$ = 4597.1 MHz, $\Delta_3/2\pi$ $\approx$ 4401.1 MHz, B = 100G and $P_0$ = 40 mW.
    }
    \label{fig:Readout_fidelity}
\end{figure}

We choose to initialize our quintet by exciting ground state atoms to $^3$P$_2$, $m_J=1$, which exhibits near-magic trapping conditions with the ground state.
In order to transfer the population from the {$^1$S$_0$}\,$\rightarrow$\,{$^3$P$_2$} we employ a multi-photon coupling scheme {$^1$S$_0$}\,$\rightarrow$\,{$^3$P$_1$}\,$\rightarrow$\,{$^3$S$_1$}\,$\rightarrow$\,{$^3$P$_2$}. Our scheme is similar to that of Ref.~\cite{ammenwerth_realization_2024,He_Coherent_three-photon,Pucher_Fine-Structure, unnikrishnan_coherent_2024}. The full scheme consists of twelve levels. To gain some intuition, we begin by modeling a minimal bare (Zeeman substructure and tweezer light shifts ignored) four-level scheme, considering only $m_J=0$ levels. We address each transition by a laser with Rabi frequency $\Omega_i$ where $i$ labels the transitions with $i=1$ corresponding to {$^1$S$_0$}\,$\rightarrow$\,{$^3$P$_1$}\,, $i=2$ to ~ {$^3$P$_1$}\,$\rightarrow$\,{$^3$S$_1$}\, and $i=3$ to{\,$^3$P$_2$}\,$\rightarrow$\,{$^3$S$_1$}. Setting the $\ket{^1\mathrm{S}_0}$ energy to zero, the Hamiltonian ($\hbar=1$) describing atom-light interaction is given by 

\begin{flalign}
    H = &-\Delta_1\ket{^3\mathrm{P}_1}\bra{^3\mathrm{P}_1} -(\Delta_1+\Delta_2)\ket{^3\mathrm{S}_1}\bra{^3\mathrm{S}_1} \\ \nonumber
    &+(\Delta_3 - \Delta_1 - \Delta_2)\ket{^3\mathrm{P}_2}\bra{^3\mathrm{P}_2} \\ \nonumber
    &+\Bigl( \frac{\Omega_1}{2} \ket{^1\mathrm{S}_0}\bra{^3\mathrm{P}_1} + \frac{\Omega_2}{2} \ket{^3\mathrm{P}_1}\bra{^3\mathrm{S}_1} \\ \nonumber
    &+ \frac{\Omega_3}{2}\ket{^3\mathrm{P}_2}\bra{^3\mathrm{S}_1} + \text{h.c.}\Bigl),
\end{flalign}
where $\Omega_i$, $\Delta_i$  are the Rabi frequencies and the detunings for each addressed transition, respectively. Including the spontaneous emission, the dynamics are governed by the Lindblad master equation

\begin{equation}
    \dot{\rho} = -i\left[ H, \rho\right] + \mathcal{D}(\rho),
\end{equation}
with
\begin{equation}
    \mathcal{D}(\rho) = \sum_{i}\left[c_i\rho c_i^{\dagger}-\frac{1}{2}\{c_i^{\dagger}c_i,\rho\}\right],
\end{equation}
where $c_i$ are the jump operators of each decay with $c_1 = \sqrt{\Gamma_1}\ket{^1\text{S}_0}\bra{^3\text{P}_1}$, $c_2 = \sqrt{\Gamma_2}\ket{^3\text{P}_1}\bra{^3\text{S}_1}$ and $c_3 = \sqrt{\Gamma_3}\ket{^3\text{P}_2}\bra{^3\text{S}_1}$ where $\Gamma_i$ are the respective decay rates. 

The effective dynamics between $^{1}$S$_0 \leftrightarrow$$ ^{3}$P$_2$ can be obtained by adiabatic elimination of the intermediate states~\cite{Reiter2012,ammenwerth_realization_2024,He_Coherent_three-photon} from which we get an effective Hamiltonian
\begin{flalign}\label{eq_eff}
    H_{\text{eff}} &= \Delta_{\text{eff}} \,\ket{^3\text{P}_2}\bra{^3\text{P}_2} \\ \nonumber &+ \frac{\Omega_{\text{eff}}}{2}\,\left(\ket{^3\text{P}_2}\bra{^1\text{S}_0} + \, \text{h.c.} \right),
\end{flalign} 
where
\begin{flalign}
    \Delta_{\text{eff}} &\simeq \Delta_3 - \Delta_1 - \Delta_2 +\frac{\Delta_1\Omega_3^2 -(\Delta_1 +\Delta_2)\Omega_1^2}{4\Delta_1(\Delta_1 + \Delta_2)-\Omega_2^2}, \\ \nonumber
    \Omega_{\text{eff}} &\simeq \frac{\Omega_1\,\Omega_2\,\Omega_3}{4\Delta_1(\Delta_1 + \Delta_2)-\Omega_2^2},
\end{flalign}
under the assumption that $\Delta_i,\Omega_i \gg \Gamma_i$. The effective operator formalism used for the adiabatic elimination~\cite{Reiter2012} assumes stable ground states and decay only from the excited states. Therefore, we can derive an approximate decay rate by eliminating $^3\text{S}_1$ and forming an effective decay from the dominant contributions as

\begin{equation}\label{decay}
    \Gamma_{\text{eff}} \simeq \frac{\Gamma_2\Omega_3^2+ \Gamma_3\Omega_2^2}{4(\Delta_1+\Delta_2)^2},
\end{equation}
which corresponds to an effective jump operator $c_{\text{eff}} = \sqrt{\Gamma_{\text{eff}}} \, \ket{^1\text{S}_0}\bra{^3\text{P}_2} $. The effective two-level transfer between $^{1}S_0 \leftrightarrow$$ ^{3}P_2$ can be achieved by selecting large detunings (in comparison to the linewidth and Rabi frequency of each transition) to minimize the population of intermediate states. Eq.\eqref{eq_eff} dictates that a resonant transition requires $\Delta_{\text{eff}} = 0$. Further tuning of the effective decay rate leads to high fidelities and guide us in selecting optimal parameters. Under real experimental conditions, it is impossible to ignore the full Zeeman substructure. Nonetheless, the effective two-level model acquired here can serve as a useful tool for optimizing the full system. 

When the Zeeman substructure is included, while each transition is driven by a single laser frequency, the detunings of individual $m_J$ states must be adjusted to account for the Zeeman and light shifts. The Rabi frequencies are appropriately modified according to the Wigner-Eckart theorem~\cite{atomic} as described in App.\ref{App:multi}. We assume each beam to have equal contributions of all polarization components. Following the principles of the effective two-level model derived from the minimal four-level system Eq.\eqref{eq_eff}, we prepare a particular $^3\text{P}_2, m_J$ state by setting individual $m_J$ detunings of the {$^1$S$_0$}\,$\leftrightarrow$\,{$^3$P$_1$}\,$\leftrightarrow$\,{$^3$S$_1$}\,$\leftrightarrow$\,{$^3$P$_2$} transition on resonance, which results in the condition
\begin{equation}\label{eq:multi_con}
    \Delta_3 = \Delta_1 + \Delta_2 + \omega_{0}^{^1\text{S}_0} - \omega_{m_J}^{^3\text{P}_2} - \Delta'_{\text{eff},}
\end{equation}
where $\omega_{m_J}^{^{2S+1}L_J}$ are the Zeeman and tweezer light shifts of the $\ket{^{2S+1}L_J , m_J}$ states. The additional term $\Delta'_{\text{eff}}$ accounts for multi-photon light shifts. Subtracting it in Eq.\eqref{eq:multi_con} ensures the effective two-level {$^1$S$_0$}\,$\leftrightarrow$\,{$^3$P$_2, m_J$} transition is on resonance and can be calculated numerically as described in App.\ref{App:multi}.

\begin{figure}[t!]
    \centering
    \includegraphics[scale=1.0]{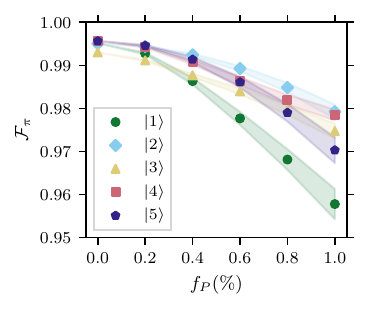}
    \caption{$\pi$-pulse fidelity $\mathcal{F}_\pi$ averaged over 100 simulations for transfer from the computational states to the ground state or vice versa within less than 1 \si{\micro\second}, depending on power fluctuations $f_P$ in both the optical tweezers and the coupling lasers. Error bands represent the standard error. Power fluctuations are accounted for each driving laser and the trapping tweezer, separately. Parameters and $\pi$ pulse duration times used can be found in Table \ref{Tab:params}.}
    \label{fig:fid_err_3}
\end{figure}

In Fig.~\ref{fig:Readout_fidelity}, we plot the dynamics of the full {$^1$S$_0$}\,$\leftrightarrow$\,{$^3$P$_2$} transition including the Zeeman substructure while satisfying the multi-photon condition Eq.~\eqref{eq:multi_con} for {$^3$P$_2, m_J=1$}. 
The qudit needs to be initialized through this state in order to avoid excessive heating through the ``squeezing effect'' caused by the trapping frequency mismatch between the {$^1$S$_0$} state and the other four qudit states. A further discussion of this effect and how to minimize heating during state preperation can be found in App.~\ref{app: Minimizing motion}.
We observe Rabi oscillations between {$^1$S$_0$}\,$\leftrightarrow$\,{$^3$P$_2, m_J=1$} where we achieve population transfer in less than 1 \si{\micro\second} with fidelity $\mathcal{F} \simeq$ 0.996. Similar transfers can be achieved for all {$^3$P$_2, m_J$} states by appropriate selection of the multi-photon condition with $\mathcal{F} \geq$ 0.99 in less than 1 \si{\micro \second }. Details of the full and effective dynamics of the transition as well as the Rabi oscillations of all {$^3$P$_2, m_J$} states can be found in App.\ref{App:multi}. The calculations are performed using QuTiP~\cite{qutip1,qutip2,qutip3}. We note that the initial state preparation, shown in Fig.~\ref{fig:pumping_schema}(a), is expected to take approximately 20~\si{\micro\second}, as it includes an additional confirmation image taken prior to the start of computation.

The $\pi$-pulse transfer fidelities $\mathcal{F}_\pi$ are affected by shot-to-shot power fluctuations in both the driving lasers and the trapping tweezer. We show this effect for each $^3\mathrm{P}_2,  m_J$ state in Fig.~\ref{fig:fid_err_3}. To simulate the effect of power fluctuations, we average the fidelities from 100 runs where the power of the trapping tweezer and the driving lasers is disturbed individually as  $P=P_{\rm{un}}(1+\beta)$ where $P_{\rm un}$ is the undisturbed power of each source, respectively, and  $\beta$ is the error drawn from a Gaussian distribution with zero mean and standard deviation $f_P$. The error $\beta$ is sampled independently for each power source. We take the standard error as a measure of the uncertainty in the fidelities. This treatment assumes the power fluctuations are on a timescale much longer than the duration of a transfer and thus remain constant during each pulse. An analytic error budget for $f_P$ = 0.2 $\%$ as well as an estimate of population leakage to the other compuational states beyond single populated state transfers can be found in App.\ref{app:multi_error_b}.

We acknowledge that the fluctuation in the power of the trapping tweezers has a more significant effect on the computed $\mathcal{F}_\pi$, as it heavily affects the multi-photon resonance condition in Eq.\eqref{eq:multi_con}, as the drive lasers power.
Transitions with a larger differential light shift, $|\omega_{0}^{^1\text{S}_0} - \omega_{m_J}^{^3\text{P}_2}|$, are more heavily affected by these fluctuations (see App.~\ref{app:Polarizability}). 
In order to reduce the sensitivity to power fluctuations, we have optimized the Rabi frequencies $\Omega_{1,2,3}$ and detunings $\Delta_{1,2,3}$ with power fluctuations present and under the constraint $\Omega_1^2/\Delta_1=\Omega_3^2/\Delta_3$, which (approximately) cancels the differential light shifts on the two states of interest~\cite{ammenwerth_realization_2024}. More details on the optimization procedure are given in App.~\ref{app:opti}.

\subsection{Computation with a RF-drive}
\label{sec:RF-drive}

\subsubsection{Computation within one quintet}
\label{sec:RF-drive_quintet}

\begin{table}[]
\caption{Tweezer parameters of each ${}^3P_2$ state in units of units of kHz  for a 1064 nm tweezer with a waist of 1 \si{\micro \meter } and a power of 40 mW.
The top table gives the scalar, vector and tensor light shifts $c_0$, $c_1$, and $c_2$ as defined in  Eq.\ \eqref{eqn:Htw}.
The bottom table gives trap depth $U_0$ and trapping frequencies in the radial (axial)  direction $\omega_{x,y}$ ($\omega_z$).
The rightmost column of the bottom table shows the differential light shifts $\Delta \rm{RF}_i$ of the RF transitions from state $i$ to $i+1$. The RF transition frequencies between those states are given by  $\rm{RF}_i / 2\pi= \Delta \rm{RF}_i +  210106  ~\mathrm{kHz}$, where the latter frequency corresponds to the Zeeman shift between neighboring states at a magnetic field of $100~\mathrm{G}$.}

\label{Tab:RFparams}
\centering
\begin{tabular}{p{0.1\linewidth} | p{0.15\linewidth} p{0.15\linewidth} p{0.15\linewidth} p{0.15\linewidth} p{0.15\linewidth} p{0.15\linewidth}} 
 \hline
 \hline
Shifts & $c_0/2 \pi$ &  $c_1/2 \pi$ &  $c_2/2 \pi$ &\\
\hline
& -20631 & -10589 &-1449& \\
 \hline
\hline
\end{tabular}
\\
\vspace{0.7cm }
\begin{tabular}{p{0.1\linewidth} | p{0.15\linewidth} p{0.15\linewidth} p{0.15\linewidth} p{0.15\linewidth} p{0.15\linewidth} p{0.15\linewidth}}
 \hline
 \hline
 State (i)&$U_0/2 \pi$ & $\omega_{x,y}/2 \pi$ & $\omega_z/2 \pi$ &$\Delta\rm{RF}_{i}/2 \pi$\\  
 \hline
\textbf{1}     & \textbf{5249}  & \textbf{49.13}  & \textbf{11.77} & -6242 \\
2     & 11491  & 72.68  & 17.41 & -9140\\ 
3     & 20631 & 97.38  & 23.32 & -12037\\ 
4     & 32668 & 122.5  & 29.34 & -14935\\ 
5     & 47603 & 147.9  & 35.42 & \\ 
 \hline
 \hline
\end{tabular}
\end{table}

After initializing the quintet in the ${}^3\mathrm{P}_2$ manifold, all single quintet manipulations will be performed by RF pulses.
The large bias magnetic field splits the Zeeman states in energy and sets the overall scale of the RF drive frequencies. The tweezer provides a tensor shift, lifting the degeneracy between the RF transition frequencies, allowing for RF pulses that individually address a single transition.
The fidelity of these pulses can be limited by several mechanisms, such as off-resonant RF driving to other quintet states, spin-motion entanglement, and dephasing by thermal motion.
In order to estimate the achievable fidelity of the RF pulses while capturing these effects, we simulate the dynamics of a single Sr atom in a 1064 nm tweezer. 

The Hamiltonian for a free quintet in a magnetic field along the $z$-axis, which is also driven by an RF field is given by $\mathcal{H}_0=\Delta_{\rm Z} \hat{J}_z+ \Omega \cos{(\omega_{\rm RF}t)} \hat{J}_x$, where $\hat{J}_i$ are spin-2 operators, $\Delta_{\rm Z}$ is the linear Zeeman splitting, $\Omega$ is the RF Rabi frequency and $\omega_{\rm RF}$ is the RF drive frequency. After transforming to the interaction picture and taking the rotating wave approximation, we obtain
\begin{equation}
    \mathcal{H}_f= \Delta_{\rm RF} \hat{J}_z+ \frac{\Omega}{2} \hat{J}_x,
    \label{eqn:H_rf}
\end{equation}
where $\Delta_{\rm RF}=\omega_{\rm RF}-\Delta_{\rm Z}$ (see App.~\ref{app:RWA}).
Next we consider the tweezer potential and the motion of the quintet, described by the Hamiltonian
\begin{equation}
    \begin{split}
   &\mathcal{H}_{\rm tw} =\sum_{i=x,y,z} \frac{\hat{p}_i^2}{2M}\\
   & +\left(c_0 \hat{1} + c_1 \hat{J}_z + c_2 \hat{J}_z^2 \right) 
   U_{\rm tw}(\hat{x},\hat{y},\hat{z}),
    \end{split}
\label{eqn:Htw}
\end{equation}
where $\hat{p}_{i=x,y,z}$ denotes the momentum along the $x,y$ or $z$ direction,  $M$ is the mass of a ${}^{88} \rm Sr$ atom, $c_0,c_1,c_2$ are the scalar, vector, and tensor light shift respectively and  $\omega_{x,y,z}$ are the reference trapping frequencies along the $x,y$ (radial) and $z$ (axial) directions. The reference trapping frequencies are given by 
\begin{equation}
\begin{split}
    \omega_{x,y}=\frac{2}{w_0}\sqrt{\frac{U_{\rm ref}}{M}}, \\
    \omega_z=\frac{
    \lambda
    }{\pi w_0^2}\sqrt{\frac{2 U_{\rm ref}}{M}},
\end{split}
\label{eqn:omega_tw}
\end{equation}
where, $U_{\rm ref}$ is the trap depth for the reference state. For the simulations in this section, $\ket{1} =\ket{^3\mathrm{P}_2, m_J=-2}$ is taken as the reference state.
Lastly, for the Gaussian tweezers considered here, the spatial profile is given by 
\begin{equation}
U_{\rm tw}(\hat{x},\hat{y},\hat{z})=\frac{1}{1+\hat{z}^2/z_0^2} \exp(-\frac{2 (\hat{x}^2+\hat{y}^2)}{w_0^2 (1+\hat{z}^2/z_0^2)}),
\label{eqn:tweezerU}
\end{equation}
where $z_0$ is the Rayleigh range.
The values of the parameters introduced in this section are given in table ~\ref{Tab:RFparams} for a $\lambda=1064$ nm tweezer with a waist of $w_0=1$ \si{\micro \meter } and a power of 40 mW, where the values for the reference state are shown in bold. The tweezer potentials have been calculated using the method outlined in~\cite{urech2023single} (see App.~\ref{app:Polarizability}).

Using $\mathcal{H}_{\rm tot}=\mathcal{H}_{\rm f}+\mathcal{H}_{\rm tw}$, which includes the Zeeman structure, we simulate the RF pulses and characterize their fidelity. Neglecting the effects of motion, we find that at a tensor shift of $|c_2| /\Omega \geq 5$, $\pi$ pulse fidelities of $F_\pi$ $ \geq 0.994$ can be achieved on all RF transitions (see App.~\ref{app:RWA}). Because the tensor shift breaks the degeneracy of the transition frequencies between adjacent $m_J$ states, the magnitude of this shift compared to the Rabi frequency puts a limit on how well the individual RF transitions can be isolated.
As a result, the same fidelities can be achieved at higher $\Omega$ as long as the tweezer power is increased such that the $c_2/\Omega$ ratio is kept constant. The RF pulse fidelity can be improved significantly by using smooth instead of square pulses. We implement a $\sin^2$ shaped pulse of length $\tau$ and pulse area $A$ given by
\begin{equation}
    \Omega(t)=\frac{4 A}{\tau}\sin^2{(\pi t/\tau)}.
\end{equation}
In order to make comparison to square pulses easier, we set $\tau$ according to the time $\tau_\pi=2\pi/\bar{\Omega}$ required for a square $\pi$ pulse with Rabi frequency $\bar{\Omega}$ and plot the fidelities as a function of the average Rabi frequency $\bar{\Omega}$ during the pulse.
For a smooth pulse of length $\tau_\pi$, the maximum required Rabi frequency (power) is higher than for an equivalent square pulse, but the Fourier spectrum is narrower.
As a result, higher average Rabi frequencies can be used without driving off-resonant transitions or excessively exciting the motion.
These smooth pulses with a higher average Rabi frequency lead to a greatly reduced sensitivity to the effect of the finite temperature of the atom and tweezer power fluctuations.
We model these pulses by Trotterizing the RF Hamiltonian in Eq.~\ref{eqn:H_rf} with 100 steps per pulse.

\begin{figure}[]
    \centering
    \includegraphics[scale=1.0]{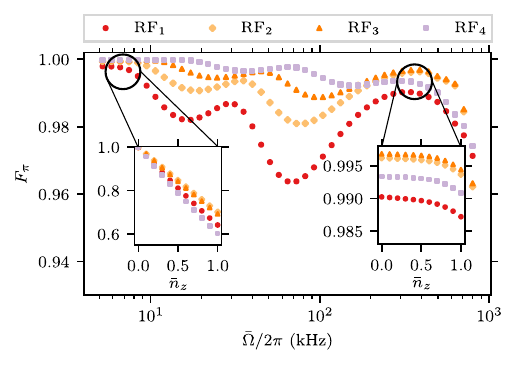}
    \caption{$\pi$ pulse fidelity as a function of the average RF Rabi frequency $\bar{\Omega}$ for each of the four RF transitions and a quintet initialized in the motional ground state. The legend indicates the transition that is being addressed, e.g. the red dots correspond to the $\rm{RF}_1$ transition. Insets: fidelity as function of the excitation number $\bar{n}_z$ for a thermal state of motion, where $\bar{n}_{x,y,z}=  \left \{ 0,0, \bar{n}_z \right \}$. Here $\bar{\Omega}/2 \pi = 6$ kHz and $\bar{\Omega}/2 \pi =  350$ kHz.} 
    \label{fig:Fidelity_Quintet}
\end{figure}
Next we investigate the effect of the RF drive strength on pulse fidelities.
For each of the four RF transitions, we initialize the quintet in the lower of the two involved states in the motional ground state.
Fig.~\ref{fig:Fidelity_Quintet} shows the $\pi$ pulse fidelities as a function of the average Rabi frequency. 
We find that for a quintet in the motional ground state, fidelities of $F_\pi$ $>0.999$ can be achieved even for square pulses at low Rabi frequencies of $\Omega/2 \pi< 10 $ kHz. 
As $\Omega$ is tuned close to one or more motional frequencies, the pulse fidelities decrease due to strong spin-motion coupling. 
At high average Rabi frequencies of order $\bar{\Omega}/2 \pi  \gg 500$ kHz, the pulse fidelities rapidly decrease as the tweezer tensor shift is no longer sufficient to isolate the RF transitions. 
In the insets of Fig.~\ref{fig:Fidelity_Quintet}, we show the impact of thermal motion in the tweezer on pulse fidelities by varying the axial excitation number $\bar{n}_z$, assuming the radial motion is ground state cooled.
The scaling of the fidelities with the radial excitation number are shown in Fig.~\ref{fig:fidelity_quintet4} in App.~\ref{app:Fid_radialmot}.
We find that the sensitivity to thermal motion increases with a lower RF Rabi frequency, and to achieve single pulse fidelities of $F\geq 0.99$, the atoms need to be cooled close to the ground state. For an operating Rabi frequency $\bar{\Omega}/2 \pi= 350$ kHz, this cooling requirement is $\bar{n}_{x,y}\leq 0.1, \bar{n}_z<0.3$.
Due to zero-point motion and the different trapping frequencies of each quintet state caused by the tensor shift needed to isolate the RF drive frequencies, even at zero temperature, the RF transition frequencies are shifted compared to the transitions in an ideal quintet trapped in a magic tweezer. As a result an additional RF detuning 
\begin{flalign}
    \Delta_{\rm ex}=\sum_{i=x,y,z}(\bar{n}_i+1/2)(\omega^k_i-\omega^l_i),
    \label{eq:squeeze_detuning}
\end{flalign} needs to be applied when driving from state $k$ to state $l$.

Using sideband cooling, the quintets can be initialized in $\bar{n}_{x,y}<0.01, \bar{n}_z \simeq 0.2$ as mentioned in Sec.~\ref{sec:Potential_experimental_realization}.
In a quintet cooled to this level and driven at $\bar{\Omega}/2 \pi= 350$ kHz, single $\pi$ pulse fidelities of $F_\pi =\{ 0.99, 0.996, 0.997, 0.993\}$ are achieved for the $\rm RF_1,RF_2,RF_3,RF_4 $ transitions respectively, with corresponding $\pi$ pulse times of $t=\{1.43,1.17,1.17,1.43 \}$ \si{\micro s}.
We note that due to the much weaker confinement along the axial direction, the effect of nonzero $\bar{n}_z$ on gate fidelities is far smaller than for similar values of $\bar{n}_{x,y}$. As shown in Fig.~\ref{fig:Fidelity_Quintet}, if the radial motion is already cooled to $\bar{n}_{x,y}=0$, cooling the axial motion further from $\bar{n}_z=1$ to $\bar{n}_z=0$ only improves the gate fidelities at $\bar{\Omega}/2\pi=350$ kHz from $F_\pi=0.983$ to $F_\pi=0.991$ for the most sensitive transition ($\rm{RF}_1$) and from $F_\pi=0.994$ to $F_\pi=0.997$ for the least sensitive transition ($\rm{RF}_3$).

Strong polarization gradients can be present in the focus of a tightly focused tweezer, leading to a state dependent deformations of the trapping potential near the focus~\cite{spreeuw2020off,Thompson_gradients}.
We investigate the impact of this effect in appendix~\ref{app:magnus}.
Taking terms which are first order in spatial derivatives into account, we find that the position of the bottom of the tweezer potential varies with the angle $\theta_z$ between the B field and tweezer polarization in a state dependent way.
These state dependent displacements excite the motion of the qudit when driving RF transitions, reducing the pulse fidelity.
However, for $\theta_z<5^\circ$, the displacement between adjacent qudit states is less than 5 nm and the effect on the RF pulse fidelities is negligible.
\\

Having analyzed the effect of thermal motion on gate fidelities, we now investigate the impact of shot-to-shot tweezer power fluctuations on the fidelity of single-quintet gates and spin-echo sequences. To model this effect, we use the same shot-to-shot power variance as introduced in Sec.~\ref{sec:state_preparation}.  

\begin{figure}[]
    \centering
    \includegraphics[scale=1]{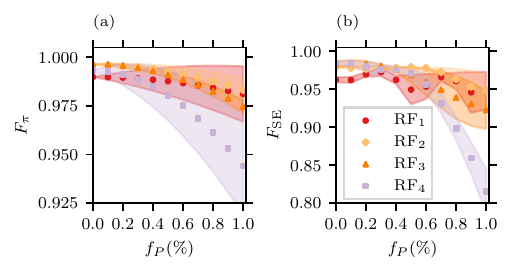}
    \caption{The effect of shot-to-shot power fluctuations for the four RF transitions on (a) $\pi$ pulse fidelity as well as (b) spin echo fidelity, each with a $\sin^2$ pulse with an average Rabi frequency of $\bar{\Omega}/2\pi= 350$ kHz, and a spin-echo dark time of 2~ms.
    The quintet was initialized in the lower of the two states of each transition and a thermal motional state with $\bar{n}_{x,y,z}=\{0.01, 0.01, 0.2 \}$. The markers correspond to the average fidelity of the two runs and the error bands correspond to the standard error.}
    \label{fig:Fidelity_powerfluct}
\end{figure}

The quintet is initialized in a thermal motional state $\bar{n}_{x,y,z} = \{0.01, 0.01, 0.2\}$. In the simulation of a $\pi$ pulse, the system is prepared in the lower of the two addressed states. For the spin-echo sequence, the system is initialized in an equal superposition of the two states, with a dark time of $2~\si{\milli s}$ (excluding the intermediate $\pi$ pulse) before a final $\pi/2$ pulse. All pulse parameters correspond to a nominal tweezer power of $P_0=40$~$ \mathrm{mW}$ and an RF Rabi frequency of $\bar{\Omega}/2 \pi =  350~\mathrm{kHz}$. Fig.~\ref{fig:Fidelity_powerfluct} shows the scaling of gate fidelity with the amplitude of power fluctuations. We clearly observe for power fluctuations below $f_{\mathrm{P}} \leq 0.2\%$, fidelities of $F\geq 0.99$ are maintained and the simple spin echo sequence is effective in mitigating the effect of shot-to-shot power fluctuations over a timescale of at least 2 ms. 
The same results give an indication of the requirements on the homogeneity of the depths of the tweezers in an array of qudits. If the depths can be balanced to the $0.3\%$ level~\cite{chew2024ultraprecise}, RF $\pi$ pulse fidelities of $F_\pi>0.988$ can be maintained over the entire array. To first order, tweezer trapping frequency inhomogeneities only affect the additional detuning in Eq.~\ref{eq:squeeze_detuning}. Assuming trap frequencies are homogenized to $5\%$ or better~\cite{chew2024ultraprecise}, the differential in shifts in Eq.~\ref{eq:squeeze_detuning} across the array are $\sim1$ kHz, resulting in a loss of fidelity for RF pulses with $\bar{\Omega}=2\pi \times 350$ kHz of less than $5\times 10^{-5}$.

\subsubsection{Single site addressability}
\label{sec:RF-drive_single_site}
In order to address a single site in the array, we keep the computational tweezers at $P_0=40$ mW and adiabatically ramp down the remaining tweezers to $P=4$ mW.
This creates a separation between the RF transition frequencies in the deep and shallow tweezers of at least $|\Delta \omega_{\rm RF}|/2 \pi=4.6 $ MHz.  
Consequently, when applying a global RF pulse resonant with the transition in the deep computational tweezers, and with a typical Rabi frequency of $\bar{\Omega}/2 \pi = 350~\mathrm{kHz}$, the populations in the shallow tweezers remain effectively unperturbed, achieving fidelities of $F\geq 0.996$. To switch the addressed site, the first computational tweezer is adiabatically ramped down while the next targeted one is ramped up over the same ramp duration. Notably, coherent driving of atoms in the shallow tweezers is still possible, albeit with a reduced Rabi frequency.

We determine the minimum adiabatic ramp time by simulating the motion during a linear ramp using the Truncated Wigner Approximation (TWA)~\cite{polkovnikov2010phase}. The quintet is initialized in state $\ket{1}$ or $\ket{5}$ and in a thermal state of motion in the deep computational tweezer. Since $\ket{1}$ and $\ket{5}$ correspond to the states with the lowest and highest trapping frequencies, respectively, they provide representative bounds for estimating the ramp-time requirements of all five quintet states. 
We find that for a ramp time of $>0.25$ ms, no appreciable heating occurs for any of the qudit states.

This ramp duration is shorter than the coherence time observed in our deep tweezers for a spin-echo sequence, and since the two ramps are symmetric, we envision applying a simple $\pi$ pulse, corresponding to the spin-echo operation, simultaneously to all states in the shallow tweezers, while computational rotations are performed in the deep tweezers, to avoid dephasing. 
In practice, the hold time $t_{\mathrm{hold}}$ will likely be reduced to the $\pi$-pulse duration in shallow tweezers, where the Rabi frequency is roughly ten times less than in deep tweezers.

\vspace{0.5cm}

With the ability to implement both local and global operations, the quintet scheme described here enables full control over all four allowed transitions in amplitude, phase, and detuning. This capability allows for the realization of arbitrary $SU(5)$ operations in the Hilbert space of each quintet. These site- and state-selective rotations, in combination with composite pulse sequences connecting quintet states that are not directly coupled, provide all the required operations for a universal single-qudit gate set~\cite{G_Brennen_Criteriaforexactqudituniversality, QuditsandHighdimensionalComputing, MultivaluedLogicCircuits}. Several proposed qudit-based schemes can be directly implemented in this architecture, including decomposition of the quintet from five states to two qubits and an ancilla state~\cite{EfficientQuditEncoding}, construction of Toffoli gates for use in Grover's algorithm~\cite{Toffoliququint}, and realization of Deutsch's algorithm within a single qudit~\cite{KIKTENKO20151409}.

\subsection{State selective readout and fast imaging}
\label{sec:SSreadout}
The state-selective readout uses the same multi-photon transition employed for state preparation, in combination with fast destructive blue imaging. When the multi-photon resonance condition in Eq.~\eqref{eq:multi_con} is satisfied, any population in the $\{^3\mathrm{P}_2, m_J\}$ manifold is transferred to the $^1\mathrm{S}_0$ state, which can then be imaged on the $^1\mathrm{S}_0\, \rightarrow\, ^1\mathrm{P}_1$ transition~\cite{Cooper_Alkaline-Earth_Atoms}.

Under typical experimental conditions, and accounting for shot-to-shot tweezer power fluctuations below $0.2\%$, all $^3$P$_2, m_J$ states can be transferred to $^1$S$_0$ with a fidelity of $\mathcal{F} > 0.99$ in less than \SI{1}{\micro s} (see Fig.~\ref{fig:fid_err_3}). The transferred population can then be imaged with high fidelity using fast ground-state imaging~\cite{Su_FAST_2025, Jackson_2020, Bluvstein_2024}, on the $\sim 10\,\si{\micro s}$ timescale, as already demonstrated in Yb tweezer arrays~\cite{Falconi_microsecondimage}. Such rapid detection is essential to fully exploit the MHz-scale Rabi frequencies of the multi-photon scheme and to minimize errors arising from quintet-state decay. Therefore, we simulate destructive imaging on the \bluetransition transition, using intensities $I \gg I_{\mathrm{sat}}$ and counter-propagating pulse trains with durations of a few hundred nanoseconds. 

By taking advantage of the fact that, at high magnetic fields, individual $m_J$ states of the \bluetransition\ transition can be spectrally resolved, and that the $^1$P$_1$, $m_J=-1$ state is strongly trapped, imaging durations of less than $10~\si{\micro s}$ could be achievable (see App.~\ref{app:Fast imaging}). 
Since the quintet states are unaffected by the blue imaging light, this fast detection scheme can be employed for mid-circuit erasure measurements to identify off-resonant scattering events~\cite{Wu2022_ErasureConversion_AlkalineEarth,Scholl2023_ErasureConversion_Rydberg}, thereby further improving the overall computational fidelity (see App.~\ref{app:Mitigating_Scattering}). 
Including additional time for atom removal from the computational array after detection, the entire quintet manifold can be mapped within $\sim100~\si{\micro s}$, which is orders of magnitude faster than any decay processes. 

\section{Conclusion and Outlook}
\label{sec:Conclusion_Outlook}
\subsection{Conclusion}
\label{sec:Conclusion}

We have presented a general method for engineering qudits through individually addressable transitions between Zeeman sublevels by combining a large linear Zeeman shift with a state-dependent light shift. As a concrete realization of this approach, we demonstrate a realistic and experimentally feasible scheme to realize and control a five-dimensional \emph{quintet} encoded in the Zeeman sublevels of the long-lived $^3$P$_2$ state of neutral $^{88}$Sr$ $ atoms confined in optical tweezers. Our modeling incorporates experimental imperfections such as shot-to-shot tweezer power fluctuations and nonzero motional occupation.   
We find that all five states of the $^3$P$_2$ \emph{quintet} can be transferred to and from the ground state with fidelities $\mathcal{F} > 0.99$ in less than \SI{1}{\micro s} using a multi-photon transfer scheme. Taking advantage of the high magnetic field and the favorable trapping conditions of the excited imaging state, we further demonstrate the feasibility of fast, destructive imaging with durations below \SI{10}{\micro s}, allowing complete state-resolved readout of the full quintet manifold within \SI{100}{\micro s}. Our analysis of single-qudit control shows that rotation fidelities $F \geq$ $0.98$ can be maintained under realistic experimental parameters taking advantage of simply easily accessible RF drive frequencies, with performance primarily limited by residual motional excitation and the stability of the tweezer intensity. By combining single-site addressability, enabled through adiabatic ramping of the individual tweezer power, more than 300 quintet rotations can be performed within \SI{100}{\milli s} while maintaining high single qudit rotation fidelity. 

\subsection{Outlook}
\label{sec:Outlook}

We plan to pursue several strategies for higher fidelities and robustness against environmental perturbations, including active atom replenishing and optimal control techniques. Here we describe how these developments, when combined with robust error mitigation, and the extension to entangling gates, will create a quintet platform useful for higher-dimensional quantum simulation, quantum sensing, and quantum computing utilizing a qudit encoded in the Zeeman sublevels of the metastable $^3$P$_2$ state in neutral $^{88}$Sr, single atom tweezer arrays.

The long RF wavelength enables coherent control over frequency, phase, and amplitude, allowing for fast and precise pulses. While our simple spin-echo scheme already improves robustness against power fluctuations, this can be extended to dynamical decoupling sequences, as widely demonstrated in qubit systems~\cite{de_Lange_Multipulse_Sensing_Sequences,Naydenov_Dynamical_decoupling}. For example, Knill pulses with interleaved delays, assembled KDD-type sequences, could further enhance superposition fidelities or tolerate larger perturbations such as tweezer power fluctuations while maintaining high fidelities~\cite{Souza_Robust_Dynamical}. Since we aim to exploit the full quintet manifold, creating global superpositions across all five states and incorporating interactions for two-quintet gates, we must also account for motional dephasing and interaction-induced errors. Here, established techniques, such as  Hamiltonian-engineered decoupling sequences offer a promising route to suppress dominant dephasing sources~\cite{Soonwon_Dynamical_Engineering,Choi_Robust_Dynamic_Hamiltonian}. In addition, recent advances in optimal pulse control, such as the Gradient Ascent Pulse Engineering (GRAPE) method using tailored phase and amplitude gradients, provide a powerful framework for designing robust, high-fidelity qudit operations~\cite{Khaneja_Optimal_control,Burshtein2025_robust}.  

In order to achieve a universal gate set, our single-quintet rotations must be supplemented by two-quintet entangling interactions. Such interactions can be engineered by coupling the $^3$P$_2$ Zeeman sublevels with a (near-)resonant drive to one of the $(5sns)$ $^3$S$_1$ Rydberg states, similar to what has been demonstrated for $^3$P$_0$~\cite{madjarov_high-fidelity_2020}. Due to the difference in $g_J$ factors for the quintet levels (2.1~MHz/G) versus the $^3$S$_1$ Rydberg states (2.8~MHz/G), unique transition frequencies for every transition are naturally available. This enables the implementation of general symmetric two-qudit entangling gates for any pair of quintet states, providing access to the full SU($d \times d$) group of two-qudit unitaries~\cite{Weggemans2021_Qudit, G_Brennen_Criteriaforexactqudituniversality}. 

Finally, the next step for the proposed computation scheme, which is not unique to our setup, is extending the computational depth despite the destructive nature of the final readout. We plan to address this in future work by combining our platform with a continuous atom source~\cite{Atom_computing,Li2025_fast_continuous_coherent,Chiu2025_continuous_3000_qubit,chen_continuous_2022,Rodrigo_MOT_2021,boughdachi2025strontium,Modeling_swadheen} or a reservoir-based approach~\cite{muniz_high-fidelity_2024,Gyger_Continuous_operation_of_large-scale}. In such schemes, a secondary, movable array of tweezers is used to restock atoms into the computational array; sorting can be realized either with crossed AODs or with a high-speed SLM~\cite{Knottnerus2025_ParallelAssembly_Tweezers_SLM,Lin2025_AIAssembly_PRL}. Continuous replenishing of lost atoms would compensate for errors due to destructive detection or preparation losses, as well as less pronounced off-resonant scattering, thereby enabling deeper quantum computations.

\begin{acknowledgments}
We acknowledge helpful discussions with Luca Guariento related to fast imaging of strontium, and Maximilian Ammenwerth and Johannes Zeiher related to multi-photon excitation. We also thank Rene Gerritsma for review and insightful comments on the manuscript.
Finally, we would like to thank referee C for suggesting to use smooth instead of square pulses, which significantly improved the quintet rotation fidelities.

This work has received funding from the European Union’s (EU) Horizon 2020 research and innovation program under Grant Agreement No. 820404 (iqClock project) and No. 860579 (MoSaiQC project). We acknowledge support from the Dutch National Growth Fund (NGF), as part of the QDNL programme. The work has also received funding under Horizon Europe programme HORIZON-CL4-2021-DIGITAL-EMERGING-01-30 via project 101070144 (EuRyQa). We thank QDNL and NWO for the grant NGF.1623.23.025 (“Qudits in theory and experiment”).
A.S.N., K.F., and B.G. were supported by the Dutch Research Council (NWO/OCW) as a part of the Quantum Software Consortium (project number 024.003.037), QDNL (project number NGF.1582.22.030) and  ENW-XL grant (project number OCENW.XL21.XL21.122).

\section*{CODE AVAILABILITY}
The data, simulation and analysis tools/code used in this manuscript can be found in Reference~\cite{3P2_2025data}. 

\end{acknowledgments}

\appendix
\section{Polarizability and scattering in the presence of the trapping tweezers}
\subsection{Polarizability in $\sigma^-$-polarized optical tweezers}
\label{app:Polarizability}
\begin{figure}[b]
    \centering
    \includegraphics[scale=1.0]{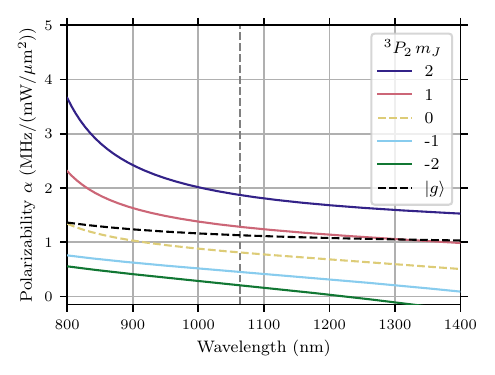}
    \caption{
    Polarizability of the different Zeeman sublevels of the $^3\mathrm{P}_2$ state and the $^1\mathrm{S}_0$ ground state as a function of wavelength for $\sigma^-$-polarized trapping light. 
    The gray dashed line indicates the wavelength of $1064~\mathrm{nm}$, used as the basis for our simulations. 
    At this wavelength, the polarizabilities (in units of MHz/(mW/\si{\micro\meter}$^2$)) are 
    $\alpha_{^3\mathrm{P}_2, m_J=-2\rightarrow+2} = 0.21,\, 0.45,\, 0.81,\, 1.28,\, 1.87$, and 
    $\alpha_{^1\mathrm{S}_0} = 1.13$.}
    \label{fig:Pol_3P2_mJ}
\end{figure}
We calculate the polarizabilities for pure $\sigma^-$-polarized trapping light following the method outlined in Ref.~\cite{urech2023single}, where scalar, vector, and tensor contributions to the light shift are included. As shown in Fig.~\ref{fig:Pol_3P2_mJ}, positive polarizabilities are obtained for all Zeeman sublevels of the $^3\mathrm{P}_2$ state as well as for the $^1\mathrm{S}_0$ ground state in the wavelength range $800$--$1150~\mathrm{nm}$. The degeneracy among the $^3\mathrm{P}_2$ sublevels is lifted due to the large tensor light shift induced by the $\sigma^-$-polarized trapping field. For simplicity, the polarizabilities are expressed in units of MHz/(mW/\si{\micro\meter}$^2$). 
The corresponding trap depth can then be written as
\begin{equation}
    U_{\mathrm{trap}} = -\alpha\, I,
    \label{eqn:trap_depth}
\end{equation}
where $I$ is the peak  intensity of the trapping beam. For a Gaussian beam this is given by
\begin{equation}
    I = \frac{2P_0}{\pi w_0^2},
\end{equation}
where $P_0$ is the power of the trapping tweezer, and $w_0$ is the waist.

We note that the wavelength region near $1064~\mathrm{nm}$ offers favorable conditions for long trapping lifetimes and efficient confinement of both ground and metastable states, while still allowing sufficiently large differential light shifts for individual \emph{quintet}-state control.

\subsection{Scattering rates due to trapping tweezers}
\label{app:Scattering}

\begin{figure}[]
    \centering
    \includegraphics[scale=1.0]{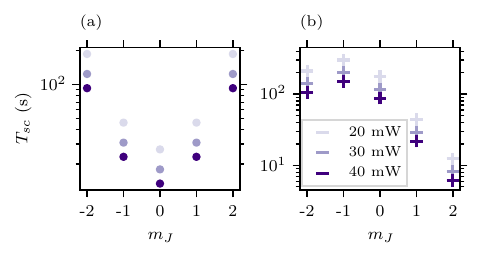}
    \caption{Lifetimes $T_{sc}$ due to off-resonant scattering caused by the trapping tweezer for each $m_J$ state of $^3\text{P}_2$  for three power values P$_0$ with (a) $\pi$ polarization and (b) $\sigma^-$ polarization.  NA = 0.5, $\lambda$ = 1064 nm,  and B = 100 G.
    }
    \label{fig:scattering_rates}
\end{figure}

The off-resonant scattering rates for a two-level system is given by~\cite{Grimm_2000}
\begin{equation}
\label{eq:Scatteringrate}
    \Gamma_{\text{sc}}= \frac{3\pi c^2}{2\hbar\omega_0^3}\left(\frac{\omega}{\omega_0}\right)^3\left(\frac{\Gamma}{\omega_0 - \omega} + \frac{\Gamma}{\omega_0 + \omega}\right)^2 I,
\end{equation}
where $\omega$ is the frequency of the tweezer, $\omega_0$ is the optical transition frequency and $\Gamma$ is the decay rate of the transition. Using Eq.\eqref{eq:Scatteringrate}, we calculate the scattering rates for each $m_J$ state of $^3\mathrm{P}_2$ by summing over the scattering rates between  $^3\mathrm{P}_2 , m_j$ and all possible measured transitions~\cite{Trautmann_2023} while accounting for the Zeeman and light shifts and appropriately adjusting the decay rates. We label the total scattering rate for each $m_J$ state as $\Gamma_{\text{tot}}$. Then, lifetimes due to scattering are defined as $T_{sc}=1/\Gamma_{\text{tot}}$ and are plotted in Fig.\ref{fig:scattering_rates} for three tweezer powers and for $\pi$ and $\sigma^-$ polarization. 

\subsection{Mitigating the effect of off-resonant scattering on computation fidelities}
\label{app:Mitigating_Scattering}

\begin{figure}[b!]
\centering
\includegraphics[scale=0.9]{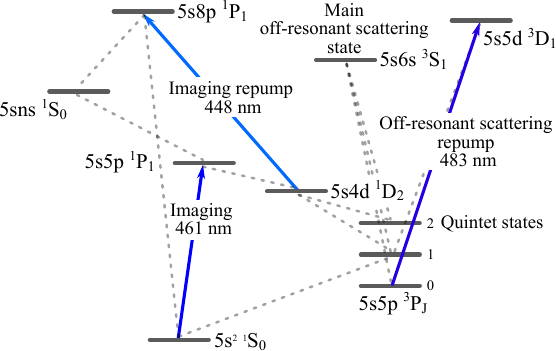}
\caption{Level diagram of $\mathrm{^{88}Sr}$ showing the main transition used for imaging, as well as two important repumping transitions (solid arrows). The relevant decay channels are indicated by dotted lines. The imaging repumpers prevent population loss to the metastable states during fluorescence detection, while the off-resonant scattering repumper enables detection and removal of atoms that have been transferred to  $^3$P$_0$ by unwanted scattering events.}
\label{fig:Repumping}
\end{figure}

An additional source of computational error arises from off-resonant scattering of the tweezer light (see fig.~\ref{fig:scattering_rates}). Although the scattering rate is very low, such processes can populate the $^3$S$_1$ state, which subsequently decays either into the long-lived metastable $^3$P$_0$ state or, via $^3$P$_1$, back to the ground state (see fig.~\ref{fig:Repumping}). Both decay channels lead to population loss. Potentially, occupation of the $^3$P$_0$ state  can reduces the detection fidelity. This effect can be mitigated by implementing a repumping scheme via the $^3$D$_1$ state using $483~\mathrm{nm}$ light, which predominantly decays through $^3$P$_1$ to the ground state~\cite{Patel2024_Sr_P0_D1, Stellmer2014_SrRepump, Okamoto2025_OpticalPumping_448nm_Sr}. Such a scheme, in combination with our fast imaging, enables direct detection of off-resonant scattering events and subtraction of their contribution, thereby minimizing their impact on computational fidelity.  

\section{Fast imaging}
\label{app:Fast imaging}

\begin{figure}[t!]
    \centering
\includegraphics[scale=1.0]{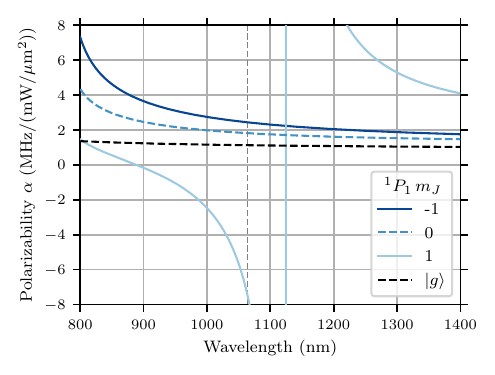}
    \caption{
    Polarizability of the different Zeeman sublevels of the $^1\mathrm{P}_1$ state and the $^1\mathrm{S}_0$ ground state as a function of wavelength for $\sigma^-$-polarized trapping light. 
    The gray dashed line indicates the wavelength of $1064~\mathrm{nm}$, used as the basis for our imaging simulations. We can clearly see that $m_J=1$ is strongly anti-trapped.
    At this wavelength, the polarizabilities (in units of MHz/(mW/\si{\micro\meter}$^2$)) are $\alpha_{^1\mathrm{P}_1,m_J=-1\rightarrow+1} = 2.44,\, 1.82,\, -7.46$, and $\alpha_{^1\mathrm{S}_0} = 1.13$.}
    \label{fig:Pol_1P1_mJ}
\end{figure}

To achieve fast imaging times, we consider counter-propagating pulse trains with durations of a few hundred nanoseconds and intensities of approximately $10\, I_{\mathrm{sat}}$. Under these conditions, the atoms experience not only the ground-state potential of $1~\mathrm{mK}$, but an effective potential given by the average over ground and excited states. With the considered magnetic field of 100 G and $\sigma^-$-polarized tweezers, the $^1$P$_1, m_J=-1$ state is trapped at 3 mK, while the $m_J=+1$ state is antitrapped at $-8$ mK. The detuning of  $\sim17\,\Gamma$ between the trapped and antitrapped state, roughly $8\,\Gamma$ from the Zeeman shift and $\sim9\,\Gamma$ from the light shift, ensuring only the trapped $m_J=-1$ state is addressed (see fig.~\ref{fig:Pol_1P1_mJ}). The trap depth is calculated based on Ref.~\cite{urech2023single}. Averaging over the time spent in both states yields an effective attractive potential of approximately 2 mK, sufficient to confine atoms in the tweezers while suppressing recoil heating and dipole-force fluctuations~\cite{Falconi_microsecondimage}.

We estimate the number of scattered photons and the atom loss rate during imaging by simulating the heating dynamics of an atom initially in the motional ground state of the tweezer~\cite{Luca_Thesis}. In our model, each absorbed photon imparts recoil in alternating directions (to represent the counterpropagating pulse sequence), while spontaneous emissions are treated as random recoils. Atoms are considered lost once their kinetic energy exceeds the trap depth. Averaging 2500 trajectories, we obtain a heating rate of $\sim 30$~\si{\micro K/\micro s} and a loss probability below $1\%$ for imaging durations of 8~\si{\micro s}, consistent with Ref.~\cite{Falconi_microsecondimage}. Including the collection efficiency of the microscope, optical losses, and a typical camera quantum efficiency at 461~nm, we estimate 5 photons/\si{\micro s} will be detected under such imaging conditions.

To detect the five $m_J$ states of the quintet as quickly as possible, we envision destructive imaging, thereby avoiding the reexcitation of ground-state atoms by the state-selective quintet deexcitation pulses. To ensure removal of imaged atoms, we include a \SI{3}{\micro s} single-sided blue pulse to heat them out of the trap, followed by \SI{6}{\micro s} for the atoms to leave the array before the next image. With 8~\si{\micro s} imaging pulses and 1~\si{\micro s} state-selective transfers, the entire quintet manifold can be mapped within 100~\si{\micro s}, orders of magnitude faster than any relevant decay process.

A complication arises because \bluetransition $\,$is not perfectly closed: atoms in the excited state have a branching ratio of about $1/50{,}000$~\cite{Hunter_Stark--electric-quadrupole, Jackson_Number-resolved, Okamoto2025DirectMeasurement} - $1/20{,}000$~\cite{Porsev_Many-body_calculations, Werij_Oscillator_strengths} to decay into the {$^1$D$_2$} state (see fig.~\ref{fig:Repumping}). From there, further decay to {$^3$P$_2$} is possible, which could lead to false detections later in the sequence and would therefore be particularly harmful to our scheme. However, this issue can be mitigated by implementing an additional repumping laser that drives the {$^1$D$_2$}\,$\rightarrow$\,{$^1$P$_1$} transition with $448~\mathrm{nm}$ light, thus preventing the population from accumulating in {$^3$P$_2$} and allowing nanosecond-scale repumping~\cite{Samland_Optical_pumping, Okamoto2025_OpticalPumping_448nm_Sr}.

With such fast repumping, the \bluetransition$\,$ can be treated as nearly closed, while still supporting scattering rates of up to tens of megahertz.  Overall, these considerations suggest that imaging with exposure times of only tens of microseconds and very high fidelity should be achievable.

\section{$^{1}S_0 \leftrightarrow$$ ^{3}P_2$ multi-photon transition dynamics}\label{App:multi}

Here we expand on the calculation of the multi-photon transition presented in the main text, including the addition of the Zeeman substructure and the full multi-photon coupling.  We additionally derive the effective $^{1}S_0 \leftrightarrow$$ \, ^{3}$P$_2 , m_J$ two-level dynamics and prove the multi-photon transition condition.

\subsection{Including Zeeman substructure}

State preparation {$^1$S$_0$}\,$\rightarrow$\,{$^3$P$_1$}\,$\rightarrow$\,{$^3$S$_1$}\,$\rightarrow$\,{$^3$P$_2$} and imaging {$^3$P$_2$} \,$\rightarrow$\,{$^3$S$_1$}\,$\rightarrow${$^3$P$_1$}\,\,$\rightarrow$\,{$^1$S$_0$} can be performed by a multi-photon coupling scheme, similarly as suggested in Ref.~\cite{ammenwerth_realization_2024,He_Coherent_three-photon,Pucher_Fine-Structure, unnikrishnan_coherent_2024}. Assuming that each transition is addressed by a single laser frequency with all polarization components and that the $\ket{^1\mathrm{S}_0}$ state is taken to be at zero energy, the Hamiltonian, including the Zeeman substructure, in a rotating frame becomes
\begin{widetext}
\begin{flalign}\label{ham_12}
    H' = &-\sum_{m_{J}}\left(\Delta_1 + \omega_0^{^1\text{S}_0} - \omega_{m_{J}}^{^3\text{P}_1}\right)\ket{^3\text{P}_1\, m_{J}}\bra{^3\text{P}_1 \, m_{J}} 
    -\sum_{m_{J}}\left(\Delta_1 + \Delta_2 + \omega_0^{^1\text{S}_0} - \omega_{m_{J}}^{^3\text{S}_1}\right)\ket{^3\text{S}_1\, m_{J}}\bra{^3\text{S}_1 \, m_{J}} \nonumber \\
    &-\sum_{m_{J}}\left(\Delta_1 + \Delta_2 - \Delta_3 + \omega_0^{^1\text{S}_0} - \omega_{m_{J}}^{^3\text{P}_2}\right)\ket{^3\text{P}_2\, m_{J}}\bra{^3\text{P}_2 \, m_{J}} \\ \nonumber
    &+\Biggr[\sum_{\substack{m_{J_g}, m_{J_e}\\ q}}\frac{1}{\sqrt{2J_e+1}}\bra{J_g m_{J_g} ; 1 q}\ket{J_e m_{J_e}} \frac{\Omega_1}{2} \ket{^3\text{P}_1\, m_{J_e}}\bra{^1\text{S}_0 \, m_{J_g}} \\ \nonumber &+\sum_{\substack{m_{J_g}, m_{J_e}\\ q}}\frac{1}{\sqrt{2J_e+1}}\bra{J_g m_{J_g} ; 1 q}\ket{J_e m_{J_e}} \frac{\Omega_2}{2} \ket{^3\text{S}_1\, m_{J_e}}\bra{^3\text{P}_1 \, m_{J_g}} \\ \nonumber &+\sum_{\substack{m_{J_g}, m_{J_e}\\ q}}\frac{1}{\sqrt{2J_e+1}}\bra{J_g m_{J_g} ; 1 q}\ket{J_e m_{J_e}} \frac{\Omega_3}{2} \ket{^3\text{S}_1\, m_{J_e}}\bra{^3\text{P}_2 \, m_{J_g}} + \text{h.c.}\Biggr],
\end{flalign}
\end{widetext}
where detunings are defined for $m_{J}=0$ states as $\Delta_{i} = \omega_{i} -(\omega_{J_e}-\omega_{J_g})$, where $\omega_i$ is the frequency of the ith laser and $\omega_{J_e}-\omega_{J_g}$ are the bare frequency differences of the addressed states. The Rabi frequencies are defined as 
\begin{equation}
    \Omega_{i} = d_{J_gJ_e}\mathcal{E}_{i}
\end{equation}
with transition dipole moments 
\begin{equation}    d_{J_gJ_e}=\bra{J_e}|e\mathbf{r}\cdot\mathbf{\hat{\epsilon}}_{q}|\ket{J_g},
\end{equation}
electric field strength of the $i^{\rm th}$ laser $\mathcal{E}_{i}$ and polarization vectors $\hat{\epsilon}_{q}$. Individual Rabi frequencies between $|J_g, m_{J_g} \rangle \leftrightarrow|J_e, m_{J_e} \rangle $ in \eqref{ham_12} are adjusted according to the Wigner-Eckart theorem~\cite{atomic}, where $\bra{J_g m_{J_g} ; 1 q}\ket{J_e m_{J_e}}$ are the Clebsch-Gordan coefficients. Lastly, $\omega_{m_J}^{^{2S+1}L_J} =g_{J}m_{J}\mu_B B/\hbar + U_{m_J}^{^{2S+1}L_J}/\hbar$ account for the Zeeman and tweezer light shifts~\cite{Cooper_Alkaline-Earth_Atoms,Safranova_2015,Trautmann_2023} of each $\ket{^{2S+1}L_J , m_J}$ state. Assuming that each beam has equal contributions of all polarizations, the atom-laser part of the Hamiltonian should be multiplied by $1/\sqrt{3}$ and our results can then be replicated by multiplying the mentioned Rabi frequencies in Table.\ref{Tab:params} by $\sqrt{3}$.

The spontaneous emission is described by 
\begin{widetext}
\begin{equation}\label{diss_multi}
    \mathcal{D'}( \rho) = \sum_{\substack{J_e \, m_{J_e}\\J_g \, m_{J_g}  }} \Gamma_{J_g m_{J_g},J_e m_{J_e}}\left[\sigma_{J_g m_{J_g},J_e m_{J_e}} \, \rho \,\sigma_{J_e m_{J_e,J_g m_{J_g}}}  
    - \frac{1}{2}\bigl\{ \sigma_{J_e m_{J_e},J_e m_{J_e}},\rho\bigl\}\right],
\end{equation}
\end{widetext}

where $\Gamma_{J_g m_{J_g},J_e m_{J_e}}$ are the decay rates between $\ket{J_e \, m_{J_e}} \rightarrow \ket{J_g \, m_{J_g}}$ and $\sigma_{J m_J,\tilde{J} m_{\tilde{J}}}$ = $|J\, m_J\rangle \langle \tilde{J} \, m_{\tilde{J}}|$. $\Gamma_{J_g m_{J_g},J_e m_{J_e}}$ are calculated using Wigner-Eckart theorem~\cite{atomic}, from which we obtain
\begin{equation}
    \Gamma_{J_g m_{J_g},J_e m_{J_e}} 
    = \frac{1}{2J_e +1}\bra{J_g m_{J_g} ; 1 q}\ket{J_e m_{J_e}}^2 \,\Gamma_{J_g,J_e},
\end{equation}
where $\Gamma_{J_g,J_e}$ are the decay rates between $\ket{J_e} \rightarrow \ket{J_g}$. The dynamics of the system are then governed by 
\begin{equation}
    \dot{\rho} = -i\left[ H', \rho\right] + \mathcal{D'}(\rho).
\end{equation}

\subsection{$^1\text{S}_0 $ $\leftrightarrow$ $^3\text{P}_2 , m_J$ multi-photon coupling}
\begin{figure*}[htpb]
\includegraphics[scale=1.0]{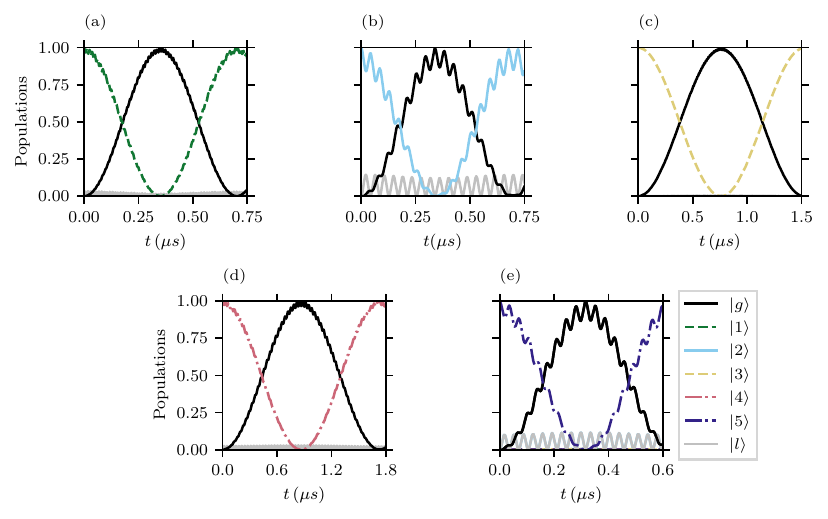}
\centering
\caption{Multi-photon coupling $^3\text{P}_2 , m_J$ $\rightarrow$ $^1\mathrm{S}_0$: Rabi oscillations between $\ket{g}$: $\ket{^1\text{S}_0}$, and (a) $\ket{1}$: $\ket{^3\text{P}_2 \, m_J=-2}$, (b) $\ket{2}$: $\ket{^3\text{P}_2 \, m_J=-1}$, (c) $\ket{3}$: $\ket{^3\text{P}_2 \, m_J=0}$, (d) $\ket{4}$: $\ket{^3\text{P}_2 \, m_J=1}$, (e) $\ket{5}$: $\ket{^3\text{P}_2 \, m_J=2}$. Fidelities $\mathcal{F}$ at $\pi$-pulse duration time $t_{\pi}$: 1: $\mathcal{F}$ = 0.995 at $t_{\pi}$ = 0.36 \si{\micro\second}, 2: $\mathcal{F}$ = 0.995 at $t_{\pi}$ = 0.34 \si{\micro\second}, 3: $\mathcal{F}$ = 0.993 at $t_{\pi}$ = 0.76 \si{\micro\second}, 4: $\mathcal{F}$ = 0.996 at $t_{\pi}$ = 0.85 \si{\micro\second}, 5: $\mathcal{F}$ = 0.996 at $t_{\pi}$ = 0.32 \si{\micro\second}. State $\ket{l}$ represents $\ket{^3\text{P}_1 \, m_J=0}$. Parameters at Table \ref{Tab:params}.}
\label{fig:imag}
\end{figure*}

The transition $^1\text{S}_0$ $\leftrightarrow$ $^3\text{P}_2 , m_J$ is achieved by tuning individual $m_J$ detunings of the {$^1$S$_0$}\,$\leftrightarrow$\,{$^3$P$_1$}\,$\leftrightarrow$\,{$^3$S$_1$}\,$\leftrightarrow$\,{$^3$P$_2$} transition on resonance, which results in the condition
$\Delta_3 = \Delta_1 + \Delta_2 + \omega_{0}^{^1\text{S}_0} - \omega_{m_J}^{^3\text{P}_2}$.  Because of the size of the system, deriving analytical equation for the effective two-level $^1\text{S}_0 \, \leftrightarrow \,^3\text{P}_2 , m_J$ system is impractical. However, we can still obtain an effective Hamiltonian numerically, using the effective operator formalism~\cite{Reiter2012}, from which we get
\begin{flalign}
    H'_{\text{eff}} &= \Delta'_{\text{eff}} \ket{^3\text{P}_2 \, m_{J}}\bra{^3\text{P}_2 \, m_{J}} \\ \nonumber &+ \frac{\Omega'_{\text{eff}}}{2}\,\Bigl(\ket{^3\text{P}_2 \,m_{J}}\bra{^1\text{S}_0 \, m_{J}=0} + \,\text{h.c} \Bigl).
\end{flalign}
To set the two-level system on resonance, we adjust the multi-photon transition condition as 
\begin{equation}\label{multi-photon_cond}
\Delta_3 = \Delta_1 + \Delta_2 + \omega_{0}^{^1\text{S}_0} - \omega_{m_J}^{^3\text{P}_2} - \Delta'_{\text{eff}}\,,
\end{equation}
guided by the effective Hamiltonian Eq.~\eqref{eq_eff}. The system can be further tuned using Eq.~\eqref{decay} to achieve optimal fidelities. State preparation/Imaging of the desired $^3\text{P}_2 , m_J$ can be then achieved by satisfying Eq.~\eqref{multi-photon_cond}. In Fig.~\ref{fig:imag} we plot the Rabi oscillations between $^1\mathrm{S}_0$ and $^3\text{P}_2$ for each $m_J$ state. The parameters used for each transition  can be found in Table~\ref{Tab:params}. The parameters are selected such that all $^3\text{P}_2 , m_J$ are prepared in less than 1 \si{\micro\second} with $ \mathcal{F} \geq $ 0.99.

\begin{table}[htpb]
\caption{Parameters used for the $^1\text{S}_0$ $\leftrightarrow$ $^3\text{P}_2 , m_J$ multi-photon coupling in MHz and $\pi$ pulse duration $t_{\pi}$ in \si{\micro\second} for B = 100G and $P_0$ = 40 mW.
All numbers have been rounded to one decimal, the exact values can be found in~\cite{3P2_2025data}.}
\centering
\begin{tabular}{p{0.09\linewidth} | p{0.1\linewidth} p{0.12\linewidth} p{0.12\linewidth} p{0.12\linewidth} p{0.12\linewidth} p{0.12\linewidth} p{0.12\linewidth} p{0.12\linewidth}} 
 \hline
 \hline
 State &$\Omega_1/2\pi$ & $\Omega_2/2\pi$ & $\Omega_3/2\pi$ & $\Delta_1/2\pi$ & $\Delta_2/2\pi$ & $\Delta_3/2\pi$ & $t_\pi$ \\  
 \hline
 $\ket{1}$ & 21 & 2717.9 & 314.5& 29.4 & 6147.4 & 6573.4 & 0.36 \\ 
 $\ket{2}$ & 13.6 & 2869.4 & 2000 & 1.9& 40673.8 &40864.1 & 0.34\\
 $\ket{3}$ & 23.2 & 4499.9 & 742.2 & 18.7 & 19095.4 & 19104.6 & 0.76\\
 $\ket{4}$ & 10.9 & 1380.6 & 219.9 & 10.8 & 4597.1  & 4401.1 & 0.85\\
 $\ket{5}$ &18.7 &2862.4 & 834.7 & 11.2 & 22555.7 &22163.9 & 0.32\\ 
 \hline
 \hline
 \label{Tab:params}
\end{tabular}
\end{table}

\subsection{Error budget}\label{app:multi_error_b}

Here, we present an analytic error budget for readout and initialization through the multi-photon transition. In Table \ref{Tab:error_multi} we provide the fidelities, residual populations, populations in ${}^3P_1$ and the rest of the states included in the multi-photon dynamics for the readout/initialization $\pi$-pulses of each computational basis, including $f_P=0.2 \%$ in power fluctuations. The residual population that remains in the initial state after readout can be mitigated by applying the multi-photon and destructive imaging pulses multiple times. Additionally, the population in ${}^3P_1$ will naturally decay to the ground state after a sufficient wait time, allowing for further improvement in fidelities.

\begin{table}[htpb]
\label{Tab:error_multi}
\caption{Error budget for readout and initialization $\pi$-pulses. `Residual' denotes the population remaining in the initial state after readout/initialization, while `Other' refers to states beyond the ground state, the computational state and ${}^3P_1$ manifold. All values are obtained assuming laser power fluctuations with standard deviation $f_P=0.2 \%$.}
\centering
\begin{tabular}{p{0.12\linewidth} | p{0.16\linewidth} p{0.16\linewidth}  p{0.16\linewidth} p{0.16\linewidth}} 
 \hline
 \hline
 State & Fidelity & Residual & ${}^3P_1$  & Other \\  
 \hline
 $\ket{1}$ & 0.9925  & 0.0029 & 0.0037 & 0.0009 \\ 
 $\ket{2}$ & 0.9943  & 0.0027 & 0.0020 &  0.0010 \\
 $\ket{3}$ & 0.9910  & 0.0007 & 0.0075 &   0.0008\\
 $\ket{4}$ & 0.9943   & 0.0010 & 0.0033 & 0.0014 \\
 $\ket{5}$ & 0.9946  & 0.0007 & 0.0044  &0.0003\\ 
 \hline
 \hline
\end{tabular}
\end{table}

Up to this point, we have considered only the case of population transfer from a single populated state. However, under realistic experimental conditions, it is required to perform imaging of computational states while other basis states exhibit non-zero population. To estimate the associated errors, we simulate population transfer from a single computational state to the ground state while the quintet is initialized in an equal superposition. For each state, we compute the total population leakage into the remaining computational states and plot it as a function of power fluctuations strength $f_P$.

\begin{figure}[htpb]
    \centering
    \includegraphics[scale=1.0]{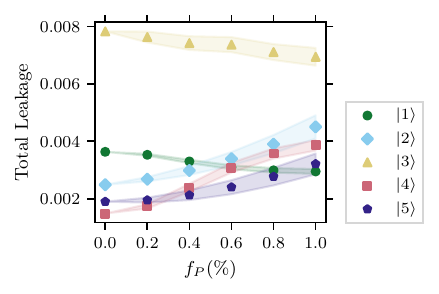}
    \caption{Total population leakage into the rest of the computational states during the transfer from $\ket{i}$ to the ground state, averaged over 100 simulations, as a function of power fluctuations $f_P$. The quintet is initialized in an equal superposition. Error bands represent the standard error. Power fluctuations are modelled independently for each driving laser and the trapping tweezer. Parameters and $\pi$ pulse duration times used can be found in Table \ref{Tab:params}.}
    \label{fig:fid_err}
\end{figure}

Optimization procedure
\label{app:opti}
We reduce the sensitivity to power fluctuations by optimizing the Rabi frequencies $\Omega_{1,2,3}$, detunings $\Delta_{1,2,3}$ and pulse time $t_\pi$ under the constraint $\Omega_1^2/\Delta_1=\Omega_3^2/\Delta_3$. During the optimization, we attempt to maximize the readout fidelity while, with equal weight, simultaneously minimizing leakage errors.
This is done by maximizing a vost function $O$ which given by the minimum of two quantities:
\begin{equation}
    O=\min \left(\tilde{F}_{g\rightarrow i}(\beta),1-\tilde{L}_{i}(\beta)\right),
\end{equation}
where $\tilde{F}_{g\rightarrow i}(\beta)=\left(F_{g\rightarrow i}(-\beta/\sqrt{2})+F_{g\rightarrow i}(\beta/\sqrt{2}) \right)/2$, $F_{g\rightarrow i}(\beta/\sqrt{2})$ is the $\ket{^{1}S_0}\rightarrow \ket{^3 P_2~i}$ $\pi$-pulse fidelity with power $P=P_0(1+\beta/\sqrt{2})$ and $\tilde{L}_{i}(\beta)=\left(L_{i}(-\beta/\sqrt{2})+L_{i}(\beta/\sqrt{2}) \right)/2$, where $L_{i}(\beta)$ is the largest fractional leakage out of the four other computational states when reading out $\ket{^3 P_2~i}$ from an equal superposition of all five qudit states with power $P=P_0(1+\beta/\sqrt{2})$.
The two power fluctuation values of $\pm\beta/\sqrt{2}$ used in the objective function were chosen to have a standard deviation of $\beta$ and zero mean.
By simultaneously optimizing over both the averaged $\bar{F}$ and $\bar{L}$, we ensure both high state preparation fidelities and low leakage errors in the presence of small power fluctuations.
The objective function is maximized using the numerical ``SimulatedAnnealing'' optimizer in Mathematica. 
The free parameters are the Rabi frequencies $\Omega_{1,2,3}$,  detunings $\Delta_{2,3}$ and pulse time $t_\pi$, while $\Delta_1=\Delta_3 \Omega_1^2 /\Omega_3^2$ in order to minimize the differential light shifts induced by the multi-photon coupling. The detuning $\Delta_3$ is constrained to a window of a few MHz around the multi-photon resonance defined in Eq.~\ref{multi-photon_cond}. We emphasize that more sophisticated optimization routines could potentially yield further improved results.

\subsection{Minimizing momentum transfer and excitation of motion}
\label{app: Minimizing motion}

As discussed in Sec.~\ref{sec:RF-drive}, achieving high-fidelity quintet rotations requires that the atoms remain very close to the motional ground state in the $^3$P$_2$ manifold. 
Since the atoms can be prepared near the motional ground state in the $^1$S$_0$ ground level, it is desirable to preserve their motional state during the initial state preparation, rather than relying on recooling once in the $^3$P$_2$ state. To this end, we must minimize the net momentum transfer from the multi-photon transition used for the initial-state transfer.
For an untrapped quintet, the net momentum transfer in the direction of $\mathbf{k}_{707\mathrm{nm}}$ can be expressed as 
\begin{equation}
\begin{split}
    |p_{in}| &= \hbar |\mathbf{k}_{\mathrm{eff}}| \\
    &= \hbar \bigl(
    \cos(\varphi_{689\mathrm{nm}})|\mathbf{k}_{689\mathrm{nm}}|
    + \cos(\varphi_{688\mathrm{nm}})|\mathbf{k}_{688\mathrm{nm}}| \\
    &\hspace{3.3em}
    - |\mathbf{k}_{707\mathrm{nm}}|
    \bigr),
\end{split}
\end{equation}
where $|\mathbf{k}_{\lambda}| = 2\pi/\lambda$ and in orthogonal direction to $\mathbf{k}_{707\mathrm{nm}}$, it can be expressed as 
\begin{equation}
\begin{split}
    |p_{ort}| &= \hbar |\mathbf{k}_{\mathrm{eff}}| \\
    &= \hbar \bigl(
    \sin(\varphi_{689\mathrm{nm}})|\mathbf{k}_{689\mathrm{nm}}|
    + \sin(\varphi_{688\mathrm{nm}})|\mathbf{k}_{688\mathrm{nm}}|),
\end{split}
\end{equation}
Those two expressions accounts for two virtual absorptions of the first two photons (first two terms) and the stimulated emission of the third~\cite{Panelli_Doppler-free_three-photon_2025,Barker_Three_photon_2016}.

We now also include the effect of the optical tweezer potential, following Ref.~\cite{Zhang2024RecoilFreeGates}. In this case, the effective momentum transfer for a $\pi$-pulse in a 2-level system is given by
\begin{equation}
    |p_R| = 
    \hbar |\mathbf{k}_{\mathrm{eff}}| 
    \frac{\Omega_{\mathrm{eff}}^2 
    |\cos\left(\frac{\pi \omega_{x,y}}{2\Omega_{\mathrm{eff}}}\right)|}{\omega_{x,y}^2},
\end{equation}
valid in the Lamb–Dicke regime ($\eta^2 (2 \bar{n}+1) \ll 1$), where 
\begin{equation}
    \eta^2 = \frac{\hbar k_{\mathrm{eff}}^2}{2M\omega_{x,y}}.
\end{equation}
Here, $\omega_{x,y}$ denotes the radial trap frequency of targeted quintet state, $M$ is the atomic mass, and $\Omega_{\mathrm{eff}}$ is the effective Rabi frequency of the multi-photon transfer. 
For this expression to be valid it is necessary for the tweezer to be magic for ${}^1\mathrm{S}_0$ state and the target state in the ${}^3\mathrm{P}_2$ manifold.

To quantify the resulting excitation of motion, we can express the mean radial motional occupation by starting in the absolute ground state as
\begin{equation}
\label{eqn:delta_n}
    \bar{n}_{x,y} = 
    \frac{\hbar |\mathbf{k}_{\mathrm{eff}}|^2 \Omega_{\mathrm{eff}}^4 
    \left[1 + \cos\left(\frac{\pi \omega_{x,y}}{2\Omega_{\mathrm{eff}}}\right)\right]}
    {4m\omega_{x,y}(\omega_{x,y}^2 - \Omega_{\mathrm{eff}}^2)^2}.
\end{equation}
However, for the fast multiphoton state preparation pulses we consider in the main text, $\Omega_{\rm{eff}}/\omega_{x,y} \gg1$ and the expression for $\bar{n}_{x,y}$ given in Eqn.~\ref{eqn:delta_n} reduces to the result for a quintet absorbing the full effective photon momentum $\hbar k_{\rm eff}$. In this limit the increase of the occupation number is given by $\bar{n}_{x,y}= \frac{\hbar k^2}{2m\omega_{x,y}}$.
Therefore, to remain close to the target value of $\bar{n}_{x,y} = 0.01$, the condition 
$|\mathbf{k}_{\mathrm{eff}}| \ll 0.3|\mathbf{k}_{707~\mathrm{nm}}|$ 
must be satisfied for $^3\mathrm{P}_2 , m_J=1$ state. 
This restriction is easily achieved experimentally by tuning the coupling beam angles close to $\varphi_{689\mathrm{nm}} = -60.9\degree$ and $\varphi_{688\mathrm{nm}} = 60.8\degree$.

When the trapping frequencies of the ${}^1\mathrm{S}_0$ ground state and the target ${}^3\mathrm{P}_2$ state are not identical, the atom can be heated not only by photon recoil but also through a \emph{squeezing effect} caused by the mismatch between the two trapping potentials. This squeezing excites breathing-mode motion of the atomic wavepacket. 

Instantaneously transferring an atom in a 1D harmonic potential with frequency $\omega_i$ to another with frequency $\omega_f$ results in a squeezed motional state characterized by a squeezing parameter $r=\frac{1}{2}\log{ \left(\omega_f/\omega_i \right)}$. For an initial thermal state with $\langle n\rangle=\bar{n}$, the corresponding motional excitation number after the transfer, is given by ~\cite{albano2002squeezed}
\begin{equation}
    \bar{n}' = \left(\bar{n}+\frac{1}{2} \right) \cosh{(2r)} -\frac{1}{2}.
\end{equation}  

To ensure that a ground-state–cooled atom remains below $\bar{n}=0.01$ after the transfer, the trapping frequencies of the ${}^1\mathrm{S}_0$ and ${}^3\mathrm{P}_2$ states must match within 22.1\%. For a tweezer wavelength of $1064\ \mathrm{nm}$, this squeezing effect increases the mean occupation from $\bar{n}=0.01$ to approximately $\bar{n}=0.011$. Accordingly, we assume that during the initialization transfer the atoms remain close to the motional ground state and therefore use the same mean occupations for the ground state and the initialized quintet state in our calculations.
Combining this requirement with the need to maintain sufficient confinement for all quintet sublevels, we identify a practical tweezer wavelength window where the above approximation holds. In particular, wavelengths in the range $850$--$1250\ \mathrm{nm}$ would enable transfer from ${}^1\mathrm{S}_0\rightarrow{}^3\mathrm{P}_2, m_J=1$ with heating below $\bar{n}=0.01$, while transfers to ${}^1\mathrm{S}_0\rightarrow{}^3\mathrm{P}_2, m_J=0$ satisfy the same condition for wavelengths near $800$--$1250\ \mathrm{nm}$.
\section{RF pulse fidelities}
\label{app: RF pulse}

\subsection{RWA validity}

\begin{figure}[b]
    \centering
    \includegraphics[scale=1]{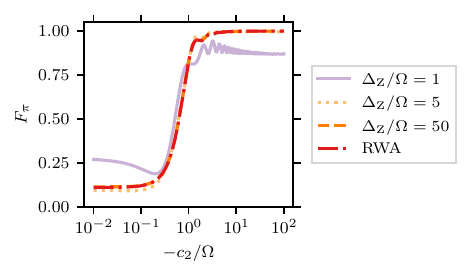}
    \caption{$\pi$ pulse fidelity on the $\ket{2} \leftrightarrow \ket{3}$ transition as a function of the tensor shift $c_2$ and Zeeman splitting $\Delta_{\rm{Z}}$, where the effects of motion have been neglected. The curves at specific values of $\Delta_Z$ correspond to the dynamics in the rotating frame without making the rotating wave approximation (RWA).}
    \label{fig:RWAplot}
\end{figure}

\label{app:RWA}
The tweezer tensor shift is required to lift the degeneracy in the RF transition frequencies in the quintet. 
As a result, the RF drive Rabi frequency is limited by the magnitude of the tensor shift $|c_2|$. 
Furthermore, at low magnetic field strength, the Rabi frequency could also by limited by the Zeeman splitting $\Delta_{\rm{Z}}$ as the rotating wave approximation (RWA) used to derive Eq.~\ref{eqn:H_rf} can break down.
Neglecting the effects of motion, we characterize the limits on $\Omega$ as a function of $\Delta_{\rm{Z}}$ and $c_2$ by simulating a $\pi$ pulse on the $\ket{2} \leftrightarrow \ket{3}$ transition using the RF Hamiltonian before the RWA was made $\mathcal{H}_0=\Delta_{\rm Z} \hat{J}_z+ \Omega \cos{(\omega_{\rm RF}t)} \hat{J}_x$ and the RWA Hamiltonian given in Eq.~\ref{eqn:H_rf}.
Due to the $\hat{J}_x$ coupling of the RF field to the quintet, the coupling between states $\ket{2},\ket{3}$ and $\ket{3}, \ket{4}$ is a factor $\sqrt{3/2}$ stronger than the coupling between $\ket{1},\ket{2}$ and $\ket{4},\ket{5}$, thus the RWA and isolation of the RF drive frequencies will first break down for the transitions coupling to $\ket{3}$.
From the curves for $F$ shown in Fig.~\ref{fig:RWAplot}, it is clear that at $\Delta_{\rm{Z}}>50\Omega$ the RWA is valid, thus at $B_{\rm{z}}=100$ the RWA can safely be applied for experimentally realizable Rabi frequencies. 
In this parameter regime, while still ignoring the motion of a quintet, $\pi$ pulse fidelities of $F_\pi \geq 0.994$ can be achieved if $|c_2/\Omega| \geq5$, giving us an upper limit on the RF Rabi frequency.

\subsection{Non-zero radial motion}
\label{app:Fid_radialmot}

\begin{figure}[]
    \centering
    \includegraphics[scale=1]{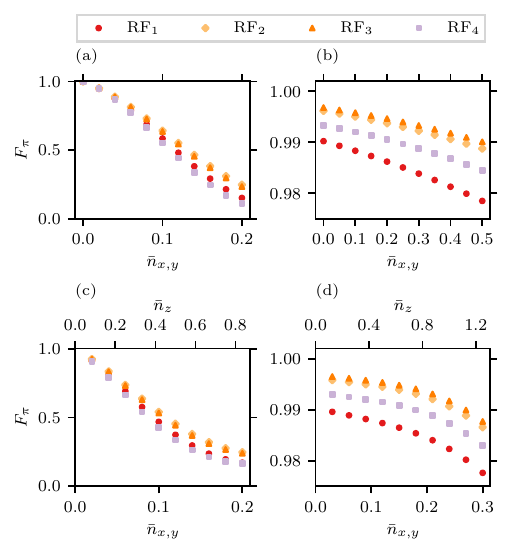}
    \caption{(a) $\pi$ pulse fidelity as function of the radial excitation numbers for $\Omega/2 \pi= 6$ kHz for the four RF transitions. 
    (b) $\pi$ pulse fidelity as function of the radial excitation numbers for $\bar{\Omega}/2 \pi= 350$ kHz.
    (c) $\pi$ pulse fidelity as function of the Temperature of a 3D motional thermal state for $\Omega/2 \pi=  48.8$$\Omega/2 \pi=6$ kHz.
     (d) $\pi$ pulse fidelity as function of the temperature of a 3D motional thermal state with corresponding occupation numbers for $\bar{\Omega}/2 \pi= 350$ kHz.}
    \label{fig:fidelity_quintet4}
\end{figure}

We further characterize the scaling of RF $\pi$ pulse fidelities with motional excitation by plotting the fidelity as a function of the radial excitation number $\bar{n}_{x,y}$ and the temperature of a 3D motional thermal state, which are the corresponding values of $\bar{n}_{x,y,z}$. 
In Fig.~\ref{fig:fidelity_quintet4}, we show the fidelities for the same two Rabi frequency values as the cross-sections in Fig.\ref{fig:Fidelity_Quintet}; $\bar{\Omega}/2 \pi= 6$ kHz and $\bar{\Omega}/2 \pi= 350$ kHz.
By comparing panels (a,b) to (b,d) and comparing to Fig. \ref{fig:Fidelity_Quintet}, it is clear that keeping the radial excitation numbers below $\bar{n}_{x,y}=0.5$ is critical to achieving fidelities $F\geq 0.98$ when driving at lower Rabi frequencies, while if $\bar{n}_{x,y}=0$, the axial excitation number can be higher than $\bar{n}_z=1$ to achieve the same fidelity.

\subsection{RF pulse error budget}
To conclude the discussion on RF pulse fidelites, we present an error budget for the RF pulses at the optimum average Rabi frequency $\bar{\Omega}=350$ kHz. In table~\ref{Tab:errorRF}, the total infidelities of $\pi$-pulses on the four RF transitions are broken down in four columns. The ``ideal'' column includes fundamental limitations like spin-motion coupling at zero temperature and off-resonant coupling to other quintet states. The ``$\bar{n}_z$'' column gives the extra infidelity caused by the atom being in a thermal state of motion with $\bar{n}_z=0.2,\bar{n}_{x,y}=0$. Similarly, the ``$\bar{n}_{x,y}$'' column gives the extra infidelity caused by the atom being in a thermal state of motion with $\bar{n}_{x,y}=0.01,\bar{n}_z=0$.
The ``$f_P$'' column gives the extra infidelity caused by shot-to-shot tweezer power fluctuations of $f_P=0.2~\% $.
Finally, the ``Total'' column gives the total pulse infidelity for an atom in a thermal state of motion with $\bar{n}_{x,y,z}=\{0.01,0.01,0.2\}$ and including tweezer power fluctuations of of $f_P=0.2~\% $.
\begin{table}[H]
\label{Tab:errorRF}
\caption{Error budget for RF $\pi$-pulses and $\bar{\Omega}/2\pi=350$ kHz.
Ideal refers to the idealized case of ground state cooled atoms and no power fluctuations.
The $\bar{n}_z$ column gives the extra error caused by a thermal occupation of $\bar{n}_z=0.2,\bar{n}_{x,y}=0$.
The $\bar{n}_{x,y}$ column gives the extra error caused by a thermal occupation of $\bar{n}_z=0,\bar{n}_{x,y}=0.01$.
The $f_P$ column gives the extra error caused by power fluctuations of $f_P=0.2 \%$.}
\centering
\begin{tabular}{p{0.2\linewidth} | p{0.13\linewidth} p{0.13\linewidth} p{0.13\linewidth} p{0.13\linewidth} p{0.13\linewidth}} 
 \hline
 \hline
 Transition &Ideal & $\bar{n}_z$ & $\bar{n}_{x,y}$ & $f_p$  & Total \\  
 \hline
 $\rm{RF}_1$ & 0.01 &$2\times 10^{-4}$  & $2\times 10^{-4}$ & $3\times 10^{-4}$  & 0.01\\ 
 $\rm{RF}_2$ &$4\times 10^{-3}$ & $1\times 10^{-4}$ & $1\times 10^{-4}$ &  $5\times 10^{-4}$& $4\times 10^{-3}$ \\
 $\rm{RF}_3$ &$3\times 10^{-3}$ & $9\times 10^{-5}$   & $8\times 10^{-5}$ & $9\times 10^{-4}$& $4\times 10^{-3}$  \\
 $\rm{RF}_4$ & $7\times 10^{-3}$ &$1\times 10^{-4}$   & $9\times 10^{-5}$  & $2\times 10^{-3}$   & $9\times 10^{-3}$  \\
 \hline
 \hline
\end{tabular}
\end{table}

\section{Adiabatic transfer between deep/shallow tweezers}
\label{app:transfer}

As stated in the main text, we characterize the tweezer depth ramp time needed to not heat the qudit during the transfer to shallower tweezers and back using the Truncated Wigner Approximation (TWA).
This phase space method can be used to exactly simulate the dynamics of a particle in an anharmonic potential by sampling from the initial Wigner quasi-probability distribution and subsequently propagating many phase space trajectories using semi-classical equations of motion~\cite{polkovnikov2010phase}. Observables such as $\langle x\rangle,\langle p\rangle$ and $\langle n\rangle$ are computed by averaging the value of the relevant Weyl symbol over many individual phase space trajectories.

Simulating the tweezer depth ramp we only consider the $\ket{1}, \ket{5}$ quintet states, as the these states experience the shallowest trapping and deepest tweezer potentials respectively and are therefore the limiting factors in the timescale of the tweezer depth ramp.
The Hamiltonian of the motion is given by Eq.~\ref{eqn:Htw}, where the tweezer shift coefficients $c_0,c_1,c_2$ are now explicitly time dependent due to the sweep of the tweezer power and are rescaled by $P(t)/P_0$, where $P(t)$ is the tweezer power at time $t$ and $P_0=40$ mW is the nominal power.
As all the functions of operators involved in the Hamiltonian are already symmetrized, we can directly get the associated Weyl symbol $H_w$ by simply replacing operators with complex numbers $\hat{a}_i \rightarrow A_i$.
Within the TWA framework, we then calculate the equations of motion for $A_i$ using the equations below
\begin{equation}
\begin{split}
&\dot{O_w}=-\frac{i}{\hbar} O_w\Lambda_c H_W,\\
   & \Lambda_c=\sum_{V= A_1,A_2,A_3}\frac{\partial^\leftarrow}{\partial V}\frac{\partial^\rightarrow}{\partial V^* }-\frac{\partial^\leftarrow}{\partial V^* }\frac{\partial^\rightarrow}{\partial V},\\
   &\dot{A}=-\frac{i}{\hbar} \frac{\partial}{\partial A^* } H_W,
\end{split}
\end{equation}
where $\partial V, \partial V^* $ correspond to partial derivatives with respect of variable V and it's complex conjugate $V^*$ respectively.
We draw the initial values of $A_i$ from Wigner functions single mode thermal states with $\langle n_\mu\rangle=\bar{n}_\mu$, which is given by
\begin{equation}
    W=\frac{1}{2 \pi \sigma_\mu^2} \exp(-\frac{\abs{A_\mu}^2}{2 \sigma_\mu^2}),
\end{equation}
where $\sigma_\mu=\sqrt{(\bar{n}_\mu+1/2)/2}$.

\begin{figure}[t!]
    \centering
    \includegraphics[scale=1]{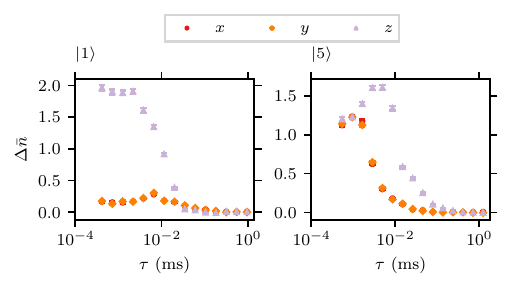}
    \caption{Increase in motional occupation numbers for the $\ket{1}$ and $\ket{5}$ qudit states after a tweezer ramp sequence with ramp length $\tau$. The tweezer power is reduced from $P=40$ mW to $P=4$ mW using a linear ramp of length $\tau$, held for 1 ms and ramped back to $P=40$ mW.}
    \label{fig:TWA_ramp_heating}
\end{figure}

During a power ramp simulation, the tweezer power is reduced from 40mW to 4 mW using a linear ramp of length $\tau$, after which the qudit is held in the shallow tweezer for $t_{\rm hold} =1$ ms, and the tweezer power is ramped up again. The qudit is initialized in a thermal motional state with $\bar{n}_{x,y,z}=\{ 0.01,0.01,0.2\}$ and the increase in $\bar{n}_{x,y,z}$ after this ramp sequence is averaged over $10^4$ trajectories. From the curves shown in Fig.~\ref{fig:TWA_ramp_heating}, we find that for $\tau>0.3$ ms no appreciable heating occurs for all qudit states.

\section{Optical Magnus effect}
\label{app:magnus}
\begin{figure}[]
    \centering
    \includegraphics[scale=1]{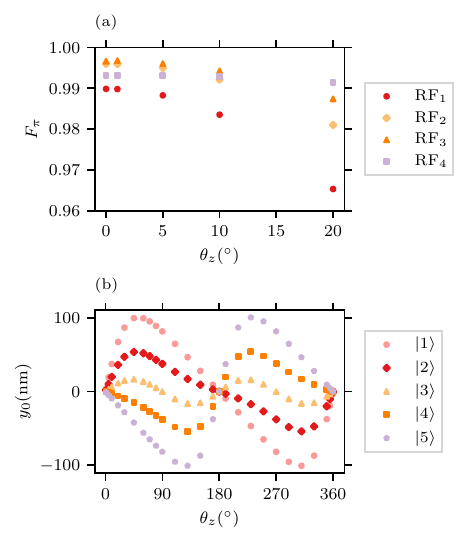}
    \caption{(a) The effect of tweezer angle errors $\theta_z$ on the fidelity of a single $\pi$ pulse for the four RF transitions. The qudit was initialised in a thermal motional state with $\bar{n}_{x,y,z}=\{0.01, 0.01, 0.2 \}$ and driven with $\Omega/2 \pi= 200$ kHz. The detuning was optimized for each tweezer angle.
    (b) Trap center positions for each qudit state as function of the tweezer anlge error different tweezer angles. The shifts of the trap centers for the different states is caused by polarization gradients in the focus of a tightly focused beam.}
    \label{fig:Magnus_shifts}
\end{figure}

To investigate the presence and impact of the optical Magnus effect~\cite{spreeuw2020off} (strong polarization gradients in the focus of a tightly focused tweezer), we use an approximate expression for the electric field profile given in~\cite{novotny2012principlesnano_optics,aiello2015transverse_angular}, which is valid up first order spatial derivatives.
This field profile is given by

\begin{equation}
    E(r,t)\simeq \rm{Re} \left[ \left(\hat{\epsilon}-i\frac{\epsilon_x x+ \epsilon_y y}{z_0} \right) f(r) e^{i (kz -\omega t)}\right],
\end{equation}
where $z_0=\pi^2 w_0^2\lambda$ and $f(r)$ is the mode function. We consider a Gaussian beam with mode function
\begin{equation}
    f(r)= \frac{w_0}{w(z)} \exp \left(-\frac{r^2}{w_0^2} \right)\exp \left(\phi(r,z) \right),
\end{equation}
where $w(z)=w_0 \sqrt{1+(z/z_0)^2}$.\\
We use this field profile to calculate the tweezer trap potential landscape around the focus of the tweezer.
For a $\pi$ polarized trap, we find that for all angles between the B-field and the tweezer $\theta_z$, the potential remains the same.
For a $\sigma_-$ polarized trap, the depths and centers of the trapping potentials for different $m_J$ states vary with the tweezer angle error $\theta_z$ as shown in panel (b) of Fig. \ref{fig:Magnus_shifts}. If the $\theta_z$ angle is kept within $\pm 10^\circ$, the displacement between the different qudit states will be less than $10$ nm. 
Because of the tensor shift needed to differentiate the RF transition frequencies, there are always large position dependent differential light shifts present in our scheme. 
In the main text and App.~\ref{app:Fid_radialmot} we have shown that when the qudit is sufficiently cooled, the infidelity caused by these differential light shifts can be minimized by using smooth RF pulses with the optimal average Rabi frequency of around $\bar{\Omega}=350$ kHz.
Any radially symmetric higher order deformations to the tweezer potential landscape caused by popularization gradients near the focus will be negligible compared to the large tensor shifts already present without taking polarization gradients into account. 
However, if large displacements between the potentials for the different Quintet states are present, the motion of the atom will be excited when driving the RF transitions, leading to lower pulse fidelities.
Approximating the potential experienced by each qudit state as a Gaussian potential centered at $y=y^i_0$ for state $\ket{i}$, we calculate the impact of tweezer angles on the fidelity of a $\pi$ pulse for a qudit initialised in a thermal motional state with $\bar{n}_{x,y,z}=\{0.01, 0.01, 0.2 \}$ and driven with $\bar{\Omega}/2 \pi= 350$ kHz.
The scaling of the fidelity with $\theta_z$ shown panel (a) of Fig. \ref{fig:Magnus_shifts} indicates that for $\theta_z<5^\circ$ the impact on $\pi$ pulse fidelities is negligible.

\bibliography{biblio_V2}

@article{ActivestabilizationofkGField,
    author = {Borkowski, Mateusz and Reichsöllner, Lukas and Thekkeppatt, Premjith and Barbé, Vincent and van Roon, Tijs and van Druten, Klaasjan and Schreck, Florian},
    title = {Active stabilization of kilogauss magnetic fields to the ppm level for magnetoassociation on ultranarrow Feshbach resonances},
    journal = {Review of Scientific Instruments},
    volume = {94},
    number = {7},
    pages = {073202},
    year = {2023},
    month = {07},
    issn = {0034-6748},
    doi = {10.1063/5.0143825},
    url = {https://doi.org/10.1063/5.0143825}
}

@article{subPPmfieldstabilityat100G,
    author = {Merkel, B. and Thirumalai, K. and Tarlton, J. E. and Schäfer, V. M. and Ballance, C. J. and Harty, T. P. and Lucas, D. M.},
    title = {Magnetic field stabilization system for atomic physics experiments},
    journal = {Review of Scientific Instruments},
    volume = {90},
    number = {4},
    pages = {044702},
    year = {2019},
    month = {04},
    issn = {0034-6748},
    doi = {10.1063/1.5080093},
    url = {https://doi.org/10.1063/1.5080093}
}

@article{
Shaw_ErasureCooling,
author = {Adam L. Shaw  and Pascal Scholl  and Ran Finkelstein  and Richard Bing-Shiun Tsai  and Joonhee Choi  and Manuel Endres },
month = {May},
title = {Erasure cooling, control, and hyperentanglement of motion in optical tweezers},
journal = {Science},
volume = {388},
number = {6749},
pages = {845-849},
year = {2025},
doi = {10.1126/science.adn2618},
URL = {https://www.science.org/doi/abs/10.1126/science.adn2618}
}

@Article{Weggemans2021_Qudit,
   author = {J. R. Weggemans and A. Urech and A. Rausch and R. Spreeuw and R. Boucherie and F. Schreck and C. J. M. Schoutens and J. Minář and F. Speelman},
   month = {June},
   title = {Solving correlation clustering with {QAOA} and a {Rydberg} qudit system: a full-stack approach},
   url = {https://quantum-journal.org/papers/q-2022-04-13-687/},
   journal = {Quantum},
   volume = {6},
   pages ={687},
   year = {2022}
}

@article{G_Brennen_Criteriaforexactqudituniversality,
  title = {Criteria for exact qudit universality},
  author = {Brennen, Gavin K. and O'Leary, Dianne P. and Bullock, Stephen S.},
  journal = {Phys. Rev. A},
  volume = {71},
  issue = {5},
  pages = {052318},
  numpages = {7},
  year = {2005},
  month = {May},
  publisher = {American Physical Society},
  doi = {10.1103/PhysRevA.71.052318},
  url = {https://link.aps.org/doi/10.1103/PhysRevA.71.052318}
}

@Article{Toffoliququint,
  author  = {A. S. Nikolaeva and E. O. Kiktenko and A. K. Fedorov},
  title   = {Generalized Toffoli Gate Decomposition Using Ququints: Towards Realizing Grover’s Algorithm with Qudits},
  journal = {Entropy},
  volume  = {25},
  pages   = {387},
  year    = {2023},
  doi     = {10.3390/e25020387},
  url     = {https://doi.org/10.3390/e25020387}
}

@article{KIKTENKO20151409,
  author  = {E. O. Kiktenko and A. K. Fedorov and A. A. Strakhov and V. I. Man'ko},
  title   = {Single qudit realization of the Deutsch algorithm using superconducting many-level quantum circuits},
  journal = {Physics Letters A},
  volume  = {379},
  pages   = {1409--1413},
  year    = {2015},
  doi     = {10.1016/j.physleta.2015.03.023},
  url     = {https://doi.org/10.1016/j.physleta.2015.03.023}
}

@article{QuditsandHighdimensionalComputing,
  author  = {Yuchen Wang and Zixuan Hu and Barry C. Sanders and Sabre Kais},
  title   = {Qudits and high-dimensional quantum computing},
  journal = {Frontiers in Physics},
  volume  = {8},
  pages   = {589504},
  year    = {2020},
  doi     = {10.3389/fphy.2020.589504},
  url     = {https://doi.org/10.3389/fphy.2020.589504}
}

@article{MultivaluedLogicCircuits,
  title = {Synthesis of multivalued quantum logic circuits by elementary gates},
  author = {Di, Yao-Min and Wei, Hai-Rui},
  journal = {Phys. Rev. A},
  volume = {87},
  issue = {1},
  pages = {012325},
  numpages = {8},
  year = {2013},
  month = {Jan},
  publisher = {American Physical Society},
  doi = {10.1103/PhysRevA.87.012325},
  url = {https://link.aps.org/doi/10.1103/PhysRevA.87.012325}
}

@article{EfficientQuditEncoding,
  author  = {A. S. Nikolaeva and E. O. Kiktenko and A. K. Fedorov},
  title   = {Efficient realization of quantum algorithms with qudits},
  journal = {EPJ Quantum Technology},
  volume  = {11},
  pages   = {43},
  year    = {2024},
  doi     = {10.1140/epjqt/s40507-024-00250-0},
  url     = {https://doi.org/10.1140/epjqt/s40507-024-00250-0}
}

@article{Schuetz_3P2_Neon,
doi = {10.1088/1361-6455/ac9c3a},
url = {https://dx.doi.org/10.1088/1361-6455/ac9c3a},
year = {2022},
month = {nov},
publisher = {IOP Publishing},
volume = {55},
number = {23},
pages = {234004},
author = {Schütz, Jan and Martin, Alexander and Laschinger, Sanah and Birkl, Gerhard},
title = {Coherent dynamics in a five-level atomic system},
journal = {Journal of Physics B: Atomic, Molecular and Optical Physics}
}

@article{Falconi_microsecondimage,
  title = {Microsecond-Scale High-Survival and Number-Resolved Detection of Ytterbium Atom Arrays},
  author = {Muzi Falconi, A. and Panza, R. and Sbernardori, S. and Forti, R. and Klemt, R. and Abdel Karim, O. and Marinelli, M. and Scazza, F.},
  journal = {Phys. Rev. Lett.},
  volume = {135},
  issue = {20},
  pages = {203402},
  numpages = {9},
  year = {2025},
  month = {Nov},
  publisher = {American Physical Society},
  doi = {10.1103/n3bg-7yw7},
  url = {https://link.aps.org/doi/10.1103/n3bg-7yw7}
}

@article{ammenwerth_realization_2024,
  title = {Realization of a Fast Triple-Magic All-Optical Qutrit in $^{88}\mathrm{Sr}$},
  author = {Ammenwerth, Maximilian and Timme, Hendrik and Gyger, Flavien and Tao, Renhao and Bloch, Immanuel and Zeiher, Johannes},
  journal = {Phys. Rev. Lett.},
  volume = {135},
  issue = {14},
  pages = {143401},
  numpages = {8},
  year = {2025},
  month = {Oct},
  publisher = {American Physical Society},
  doi = {10.1103/4kdl-xcwz},
  url = {https://link.aps.org/doi/10.1103/4kdl-xcwz}
}

@book{Luca_Thesis,
	title={A new setup for single Strontium
atoms in Optical Tweezers},
	author={Luca Guariento},
	year={2025},
	publisher={Universit\'a di Firenze}
}

@article{omanakuttan_quantum_2021,
	title = {Quantum optimal control of ten-level nuclear spin qudits in {Sr} 87},
	volume = {104},
	issn = {2469-9926, 2469-9934},
	url = {https://link.aps.org/doi/10.1103/PhysRevA.104.L060401},
	doi = {10.1103/PhysRevA.104.L060401},
	number = {6},
	journal = {Physical Review A},
	author = {Omanakuttan, Sivaprasad and Mitra, Anupam and Martin, Michael J. and Deutsch, Ivan H.},
	month = {dec},
	year = {2021},
	pages = {L060401},
}

@article{omanakuttan_qudit_2023,
    title = {Qudit Entanglers Using Quantum Optimal Control},
    author = {Omanakuttan, Sivaprasad and Mitra, Anupam and Meier, Eric J. and Martin, Michael J. and Deutsch, Ivan H.},
    journal = {PRX Quantum},
    volume = {4},
    issue = {4},
    pages = {040333},
    numpages = {16},
    year = {2023},
    month = {Nov},
    publisher = {American Physical Society},
    doi = {10.1103/PRXQuantum.4.040333},
    url = {https://link.aps.org/doi/10.1103/PRXQuantum.4.040333}
}

@article{lindon_complete_2023,
	title = {Complete {Unitary} {Qutrit} {Control} in {Ultracold} {Atoms}},
	volume = {19},
	issn = {2331-7019},
	url = {https://link.aps.org/doi/10.1103/PhysRevApplied.19.034089},
	doi = {10.1103/PhysRevApplied.19.034089},
	number = {3},
	urldate = {2025-03-26},
	journal = {Physical Review Applied},
	author = {Lindon, Joseph and Tashchilina, Arina and Cooke, Logan W. and LeBlanc, Lindsay J.},
	month = mar,
	year = {2023},
	pages = {034089},
}

@article{nakamura_hybrid_2024,
	title = {Hybrid Atom Tweezer Array of Nuclear Spin and Optical Clock Qubits},
  author = {Nakamura, Yuma and Kusano, Toshi and Yokoyama, Rei and Saito, Keito and Higashi, Koichiro and Ozawa, Naoya and Takano, Tetsushi and Takasu, Yosuke and Takahashi, Yoshiro},
  journal = {Phys. Rev. X},
  volume = {14},
  issue = {4},
  pages = {041062},
  numpages = {14},
  year = {2024},
  month = {Dec},
  publisher = {American Physical Society},
  doi = {10.1103/PhysRevX.14.041062},
  url = {https://link.aps.org/doi/10.1103/PhysRevX.14.041062}
}

@article{jia_architecture_2024,
  author  = {Zhubing Jia and William Huie and Lintao Li and Won Kyu Calvin Sun and Xiye Hu and Aakash and Healey Kogan and Abhishek Karve and Jong Yeon Lee and Jacob P. Covey and et al.},
  title   = {An architecture for two-qubit encoding in neutral ytterbium-171 atoms},
  journal = {npj Quantum Information},
  volume  = {10},
  pages   = {106},
  year    = {2024},
  doi     = {10.1038/s41534-024-00898-7},
  url     = {https://doi.org/10.1038/s41534-024-00898-7}
}

@article{reichardt_logical_2024,
  author  = {Ben W. Reichardt and Adam Paetznick and David Aasen and Ivan Basov and Juan M. Bello-Rivas and Parsa Bonderson and Rui Chao and Wim van Dam and Matthew B. Hastings and Andres Paz and et al.},
  title   = {Logical computation demonstrated with a neutral atom quantum processor},
  journal = {arXiv preprint},
  volume  = {arXiv:2411.11822},
  year    = {2024},
  url     = {https://doi.org/10.48550/arXiv.2411.11822}
}

@Article{holman_trapping_2024,
author={Holman, Aaron
and Xu, Yuan
and Sun, Ximo
and Wu, Jiahao
and Wang, Mingxuan
and Zhu, Zezheng
and Seo, Bojeong
and Yu, Nanfang
and Will, Sebastian},
title={Trapping of single atoms in metasurface optical tweezer arrays},
journal={Nature},
year={2026},
month={Jan},
day={01},
volume={649},
number={8098},
pages={859-865},
issn={1476-4687},
doi={10.1038/s41586-025-09961-5},
url={https://doi.org/10.1038/s41586-025-09961-5}
}

@article{He_Coherent_three-photon,
  title = {Coherent three-photon excitation of the strontium clock transition},
  author = {He, Junyu and Pasquiou, Benjamin and Escudero, Rodrigo Gonz\'alez and Zhou, Sheng and Borkowski, Mateusz and Schreck, Florian},
  journal = {Phys. Rev. Res.},
  volume = {7},
  issue = {1},
  pages = {L012050},
  numpages = {8},
  year = {2025},
  month = {Mar},
  publisher = {American Physical Society},
  doi = {10.1103/PhysRevResearch.7.L012050},
  url = {https://link.aps.org/doi/10.1103/PhysRevResearch.7.L012050}
}

@article{Pucher_Fine-Structure,
  title = {Fine-Structure Qubit Encoded in Metastable Strontium Trapped in an Optical Lattice},
  author = {Pucher, S. and Kl\"usener, V. and Spriestersbach, F. and Geiger, J. and Schindewolf, A. and Bloch, I. and Blatt, S.},
  journal = {Phys. Rev. Lett.},
  volume = {132},
  issue = {15},
  pages = {150605},
  numpages = {6},
  year = {2024},
  month = {Apr},
  publisher = {American Physical Society},
  doi = {10.1103/PhysRevLett.132.150605},
  url = {https://link.aps.org/doi/10.1103/PhysRevLett.132.150605}
}

@article{unnikrishnan_coherent_2024,
	title = {Coherent Control of the Fine-Structure Qubit in a Single Alkaline-Earth Atom},
  author = {Unnikrishnan, G. and Ilzh\"ofer, P. and Scholz, A. and H\"olzl, C. and G\"otzelmann, A. and Gupta, R. K. and Zhao, J. and Krauter, J. and Weber, S. and Makki, N. and B\"uchler, H. P. and Pfau, T. and Meinert, F.},
  journal = {Phys. Rev. Lett.},
  volume = {132},
  issue = {15},
  pages = {150606},
  numpages = {6},
  year = {2024},
  month = {Apr},
  publisher = {American Physical Society},
  doi = {10.1103/PhysRevLett.132.150606},
  url = {https://link.aps.org/doi/10.1103/PhysRevLett.132.150606}
}

@article{madjarov_high-fidelity_2020,
	title = {High-fidelity entanglement and detection of alkaline-earth {Rydberg} atoms},
	volume = {16},
	issn = {1745-2473, 1745-2481},
	url = {https://www.nature.com/articles/s41567-020-0903-z},
	doi = {10.1038/s41567-020-0903-z},
	number = {8},
	urldate = {2025-03-26},
	journal = {Nature Physics},
	author = {Madjarov, Ivaylo S. and Covey, Jacob P. and Shaw, Adam L. and Choi, Joonhee and Kale, Anant and Cooper, Alexandre and Pichler, Hannes and Schkolnik, Vladimir and Williams, Jason R. and Endres, Manuel},
	month = aug,
	year = {2020},
	pages = {857--861},
}

@article{tao_high-fidelity_2024,
 title = {High-Fidelity Detection of Large-Scale Atom Arrays in an Optical Lattice},
  author = {Tao, Renhao and Ammenwerth, Maximilian and Gyger, Flavien and Bloch, Immanuel and Zeiher, Johannes},
  journal = {Phys. Rev. Lett.},
  volume = {133},
  issue = {1},
  pages = {013401},
  numpages = {7},
  year = {2024},
  month = {Jul},
  publisher = {American Physical Society},
  doi = {10.1103/PhysRevLett.133.013401},
  url = {https://link.aps.org/doi/10.1103/PhysRevLett.133.013401}
}

@article{lis_mid-circuit_2023,
  title = {Midcircuit Operations Using the omg Architecture in Neutral Atom Arrays},
  author = {Lis, Joanna W. and Senoo, Aruku and McGrew, William F. and R\"onchen, Felix and Jenkins, Alec and Kaufman, Adam M.},
  journal = {Phys. Rev. X},
  volume = {13},
  number = {4},
  pages = {041035},
  year = {2023},
  month = {Nov},
  doi = {10.1103/PhysRevX.13.041035},
  url = {https://link.aps.org/doi/10.1103/PhysRevX.13.041035}
}

@article{young_half-minute-scale_2020,
	title = {Half-minute-scale atomic coherence and high relative stability in a tweezer clock},
	volume = {588},
	issn = {0028-0836, 1476-4687},
	url = {https://www.nature.com/articles/s41586-020-3009-y},
	doi = {10.1038/s41586-020-3009-y},
	number = {7838},
	urldate = {2025-03-26},
	journal = {Nature},
	author = {Young, Aaron W. and Eckner, William J. and Milner, William R. and Kedar, Dhruv and Norcia, Matthew A. and Oelker, Eric and Schine, Nathan and Ye, Jun and Kaufman, Adam M.},
	month = dec,
	year = {2020},
	pages = {408--413},
}

@article{jenkins_ytterbium_2022,
	title = {Ytterbium {Nuclear}-{Spin} {Qubits} in an {Optical} {Tweezer} {Array}},
	volume = {12},
	issn = {2160-3308},
	url = {https://link.aps.org/doi/10.1103/PhysRevX.12.021027},
	doi = {10.1103/PhysRevX.12.021027},
	number = {2},
	urldate = {2025-03-26},
	journal = {Physical Review X},
	author = {Jenkins, Alec and Lis, Joanna W. and Senoo, Aruku and McGrew, William F. and Kaufman, Adam M.},
	month = may,
	year = {2022},
	pages = {021027},
}

@article{eckner_realizing_2023,
	title = {Realizing spin squeezing with {Rydberg} interactions in an optical clock},
	volume = {621},
	issn = {0028-0836, 1476-4687},
	url = {https://www.nature.com/articles/s41586-023-06360-6},
	doi = {10.1038/s41586-023-06360-6},
	number = {7980},
	urldate = {2025-03-26},
	journal = {Nature},
	author = {Eckner, William J. and Darkwah Oppong, Nelson and Cao, Alec and Young, Aaron W. and Milner, William R. and Robinson, John M. and Ye, Jun and Kaufman, Adam M.},
	month = sep,
	year = {2023},
	pages = {734--739},
}

@article{cao_multi-qubit_2024,
	title = {Multi-qubit gates and {Schrödinger} cat states in an optical clock},
	volume = {634},
	issn = {0028-0836, 1476-4687},
	url = {https://www.nature.com/articles/s41586-024-07913-z},
	doi = {10.1038/s41586-024-07913-z},
	number = {8033},
	urldate = {2025-03-26},
	journal = {Nature},
	author = {Cao, Alec and Eckner, William J. and Lukin Yelin, Theodor and Young, Aaron W. and Jandura, Sven and Yan, Lingfeng and Kim, Kyungtae and Pupillo, Guido and Ye, Jun and Darkwah Oppong, Nelson and Kaufman, Adam M.},
	month = oct,
	year = {2024},
	pages = {315--320},
}

@article{Gyger_Continuous_operation_of_large-scale,
  title = {Continuous operation of large-scale atom arrays in optical lattices},
  author = {Gyger, Flavien and Ammenwerth, Maximilian and Tao, Renhao and Timme, Hendrik and Snigirev, Stepan and Bloch, Immanuel and Zeiher, Johannes},
  journal = {Phys. Rev. Res.},
  volume = {6},
  issue = {3},
  pages = {033104},
  numpages = {9},
  year = {2024},
  month = {Jul},
  publisher = {American Physical Society},
  doi = {10.1103/PhysRevResearch.6.033104},
  url = {https://link.aps.org/doi/10.1103/PhysRevResearch.6.033104}
}

@article{chen_continuous_2022,
	title = {Continuous {Bose}–{Einstein} condensation},
	volume = {606},
	issn = {0028-0836, 1476-4687},
	url = {https://www.nature.com/articles/s41586-022-04731-z},
	doi = {10.1038/s41586-022-04731-z},
	number = {7915},
	urldate = {2025-03-26},
	journal = {Nature},
	author = {Chen, Chun-Chia and González Escudero, Rodrigo and Minář, Jiří and Pasquiou, Benjamin and Bennetts, Shayne and Schreck, Florian},
	month = jun,
	year = {2022},
	pages = {683--687},
}

@article{Cooper_Alkaline-Earth_Atoms,
  title = {Alkaline-Earth Atoms in Optical Tweezers},
  author = {Cooper, Alexandre and Covey, Jacob P. and Madjarov, Ivaylo S. and Porsev, Sergey G. and Safronova, Marianna S. and Endres, Manuel},
  journal = {Phys. Rev. X},
  volume = {8},
  issue = {4},
  pages = {041055},
  numpages = {19},
  year = {2018},
  month = {Dec},
  publisher = {American Physical Society},
  doi = {10.1103/PhysRevX.8.041055},
  url = {https://link.aps.org/doi/10.1103/PhysRevX.8.041055}
}

@article{Samland_Optical_pumping,
  title = {Optical pumping of $5s4{d}^{1}{D}_{2}$ strontium atoms for laser cooling and imaging},
  author = {Samland, Jens and Bennetts, Shayne and Chen, Chun-Chia and Escudero, Rodrigo Gonz\'alez and Schreck, Florian and Pasquiou, Benjamin},
  journal = {Phys. Rev. Res.},
  volume = {6},
  issue = {1},
  pages = {013319},
  numpages = {10},
  year = {2024},
  month = {Mar},
  publisher = {American Physical Society},
  doi = {10.1103/PhysRevResearch.6.013319},
  url = {https://link.aps.org/doi/10.1103/PhysRevResearch.6.013319}
}

@Article{Jackson_Number-resolved,
	title={{Number-resolved imaging of $^{88}$Sr atoms in a long working distance optical tweezer}},
	author={Niamh Christina Jackson and Ryan Keith Hanley and Matthew Hill and Frédéric Leroux and Charles S. Adams and Matthew Philip Austin Jones},
	journal={SciPost Phys.},
	volume={8},
	pages={038},
	year={2020},
	publisher={SciPost},
	doi={10.21468/SciPostPhys.8.3.038},
	url={https://scipost.org/10.21468/SciPostPhys.8.3.038},
}

@article{Hunter_Stark--electric-quadrupole,
  title = {Observation of an atomic Stark--electric-quadrupole interference},
  author = {Hunter, L. R. and Walker, W. A. and Weiss, D. S.},
  journal = {Phys. Rev. Lett.},
  volume = {56},
  issue = {8},
  pages = {823--826},
  numpages = {0},
  year = {1986},
  month = {Feb},
  publisher = {American Physical Society},
  doi = {10.1103/PhysRevLett.56.823},
  url = {https://link.aps.org/doi/10.1103/PhysRevLett.56.823}
}

@article{Porsev_Many-body_calculations,
  title = {Many-body calculations of electric-dipole amplitudes for transitions between low-lying levels of Mg, Ca, and Sr},
  author = {Porsev, S. G. and Kozlov, M. G. and Rakhlina, Yu. G. and Derevianko, A.},
  journal = {Phys. Rev. A},
  volume = {64},
  issue = {1},
  pages = {012508},
  numpages = {7},
  year = {2001},
  month = {Jun},
  publisher = {American Physical Society},
  doi = {10.1103/PhysRevA.64.012508},
  url = {https://link.aps.org/doi/10.1103/PhysRevA.64.012508}
}

@article{Werij_Oscillator_strengths,
  title = {Oscillator strengths and radiative branching ratios in atomic Sr},
  author = {Werij, H. G. C. and Greene, Chris H. and Theodosiou, C. E. and Gallagher, Alan},
  journal = {Phys. Rev. A},
  volume = {46},
  issue = {3},
  pages = {1248--1260},
  numpages = {0},
  year = {1992},
  month = {Aug},
  publisher = {American Physical Society},
  doi = {10.1103/PhysRevA.46.1248},
  url = {https://link.aps.org/doi/10.1103/PhysRevA.46.1248}
}

@article{Shaw_Dark-State_Enhanced,
  title = {Dark-State Enhanced Loading of an Optical Tweezer Array},
  author = {Shaw, Adam L. and Scholl, Pascal and Finklestein, Ran and Madjarov, Ivaylo S. and Grinkemeyer, Brandon and Endres, Manuel},
  journal = {Phys. Rev. Lett.},
  volume = {130},
  issue = {19},
  pages = {193402},
  numpages = {6},
  year = {2023},
  month = {May},
  publisher = {American Physical Society},
  doi = {10.1103/PhysRevLett.130.193402},
  url = {https://link.aps.org/doi/10.1103/PhysRevLett.130.193402}
}

@article{sonderhouse_thermodynamics_2020,
	title = {Thermodynamics of a deeply degenerate {SU}({N})-symmetric {Fermi} gas},
	volume = {16},
	issn = {1745-2473, 1745-2481},
	url = {https://www.nature.com/articles/s41567-020-0986-6},
	doi = {10.1038/s41567-020-0986-6},
	number = {12},
	urldate = {2025-03-26},
	journal = {Nature Physics},
	author = {Sonderhouse, Lindsay and Sanner, Christian and Hutson, Ross B. and Goban, Akihisa and Bilitewski, Thomas and Yan, Lingfeng and Milner, William R. and Rey, Ana M. and Ye, Jun},
	month = dec,
	year = {2020},
	pages = {1216--1221},
}

@article{Soonwon_Dynamical_Engineering,
  title = {Dynamical Engineering of Interactions in Qudit Ensembles},
  author = {Choi, Soonwon and Yao, Norman Y. and Lukin, Mikhail D.},
  journal = {Phys. Rev. Lett.},
  volume = {119},
  issue = {18},
  pages = {183603},
  numpages = {6},
  year = {2017},
  month = {Nov},
  publisher = {American Physical Society},
  doi = {10.1103/PhysRevLett.119.183603},
  url = {https://link.aps.org/doi/10.1103/PhysRevLett.119.183603}
}

@article{Souza_Robust_Dynamical,
  title = {Robust Dynamical Decoupling for Quantum Computing and Quantum Memory},
  author = {Souza, Alexandre M. and \'Alvarez, Gonzalo A. and Suter, Dieter},
  journal = {Phys. Rev. Lett.},
  volume = {106},
  issue = {24},
  pages = {240501},
  numpages = {4},
  year = {2011},
  month = {Jun},
  publisher = {American Physical Society},
  doi = {10.1103/PhysRevLett.106.240501},
  url = {https://link.aps.org/doi/10.1103/PhysRevLett.106.240501}
}

@article{Choi_Robust_Dynamic_Hamiltonian,
  title = {Robust Dynamic Hamiltonian Engineering of Many-Body Spin Systems},
  author = {Choi, Joonhee and Zhou, Hengyun and Knowles, Helena S. and Landig, Renate and Choi, Soonwon and Lukin, Mikhail D.},
  journal = {Phys. Rev. X},
  volume = {10},
  issue = {3},
  pages = {031002},
  numpages = {27},
  year = {2020},
  month = {Jul},
  publisher = {American Physical Society},
  doi = {10.1103/PhysRevX.10.031002},
  url = {https://link.aps.org/doi/10.1103/PhysRevX.10.031002}
}

@article{Nataf_Exact_Diagonalization,
  title = {Exact Diagonalization of Heisenberg $\mathrm{SU}(N)$ Models},
  author = {Nataf, Pierre and Mila, Fr\'ed\'eric},
  journal = {Phys. Rev. Lett.},
  volume = {113},
  issue = {12},
  pages = {127204},
  numpages = {5},
  year = {2014},
  month = {Sep},
  publisher = {American Physical Society},
  doi = {10.1103/PhysRevLett.113.127204},
  url = {https://link.aps.org/doi/10.1103/PhysRevLett.113.127204}
}

@article{Yamamoto_Quantum_and_Thermal_Phase,
  title = {Quantum and Thermal Phase Transitions of the Triangular SU(3) Heisenberg Model under Magnetic Fields},
  author = {Yamamoto, Daisuke and Suzuki, Chihiro and Marmorini, Giacomo and Okazaki, Sho and Furukawa, Nobuo},
  journal = {Phys. Rev. Lett.},
  volume = {125},
  issue = {5},
  pages = {057204},
  numpages = {6},
  year = {2020},
  month = {Jul},
  publisher = {American Physical Society},
  doi = {10.1103/PhysRevLett.125.057204},
  url = {https://link.aps.org/doi/10.1103/PhysRevLett.125.057204}
}

@article{Urech_Narrow-line_imaging,
  title = {Narrow-line imaging of single strontium atoms in shallow optical tweezers},
  author = {Urech, Alexander and Knottnerus, Ivo H. A. and Spreeuw, Robert J. C. and Schreck, Florian},
  journal = {Phys. Rev. Res.},
  volume = {4},
  issue = {2},
  pages = {023245},
  numpages = {11},
  year = {2022},
  month = {Jun},
  publisher = {American Physical Society},
  doi = {10.1103/PhysRevResearch.4.023245},
  url = {https://link.aps.org/doi/10.1103/PhysRevResearch.4.023245}
}

@book{novotny2012principlesnano_optics,
  author = {Novotny, L. and Hecht, B.},
  title = {Principles of Nano-Optics},
  year = {2012},
  publisher = {Cambridge University Press},
  url = {https://doi.org/10.1017/CBO9780511794193},
  isbn = {978-0521834210}
}

@article{aiello2015transverse_angular,
  author = {Aiello, A. and Banzer, P. and Neugebauer, M. and Leuchs, G.},
  title = {From transverse angular momentum to photonic wheels},
  journal = {Nat. Photonics},
  volume = {9},
  number = {12},
  pages = {789--795},
  year = {2015},
  doi = {10.1038/nphoton.2015.203},
  url = {https://www.nature.com/articles/nphoton.2015.203}
}

@article{Reiter2012,
  title = {Effective operator formalism for open quantum systems},
  author = {Reiter, Florentin and S\o{}rensen, Anders S.},
  journal = {Phys. Rev. A},
  volume = {85},
  issue = {3},
  pages = {032111},
  numpages = {11},
  year = {2012},
  month = {Mar},
  publisher = {American Physical Society},
  doi = {10.1103/PhysRevA.85.032111},
  url = {https://link.aps.org/doi/10.1103/PhysRevA.85.032111}
}

@book{Grimm_2000,
  author = {Grimm, R. and Weidemüller, M. and Ovchinnikov, Y. B.},
  title = {Optical Dipole Traps for Neutral Atoms},
  editor = {Bederson, B. and Walther, H.},
  series = {Advances in Atomic, Molecular, and Optical Physics},
  volume = {42},
  pages = {95--170},
  year = {2000},
  publisher = {Academic Press},
  doi = {10.1016/S1049-250X(08)60186-X},
  url = {https://www.sciencedirect.com/science/article/pii/S1049250X0860186X},
  issn = {1049-250X}
}

@article{Safranova_2015,
  title = {Extracting transition rates from zero-polarizability spectroscopy},
  author = {Safronova, M. S. and Zuhrianda, Z. and Safronova, U. I. and Clark, Charles W.},
  journal = {Phys. Rev. A},
  volume = {92},
  issue = {4},
  pages = {040501},
  numpages = {5},
  year = {2015},
  month = {Oct},
  publisher = {American Physical Society},
  doi = {10.1103/PhysRevA.92.040501},
  url = {https://link.aps.org/doi/10.1103/PhysRevA.92.040501}
}

@article{Trautmann_2023,
  title = {$^{1}\mathrm{S}_{0}\text{\ensuremath{-}}^{3}\mathrm{P}_{2}$ magnetic quadrupole transition in neutral strontium},
  author = {Trautmann, J. and Yankelev, D. and Kl\"usener, V. and Park, A. J. and Bloch, I. and Blatt, S.},
  journal = {Phys. Rev. Res.},
  volume = {5},
  issue = {1},
  pages = {013219},
  numpages = {20},
  year = {2023},
  month = {Mar},
  publisher = {American Physical Society},
  doi = {10.1103/PhysRevResearch.5.013219},
  url = {https://link.aps.org/doi/10.1103/PhysRevResearch.5.013219}
}

@article{Patel2024_Sr_P0_D1,
  author       = {Kushal Patel and Palki Gakkhar and Korak Biswas and S.\ Sagar Maurya and Pranab Dutta and Vishal Lal and B.\ K.\ Mani and Umakant D.\ Rapol},
  title        = {Spectroscopy of the 5s5p {$^3P_0$} → 5s5d {$^3D_1$} transition of strontium using laser cooled atoms},
  journal      = {Journal of Physics B: Atomic, Molecular and Optical Physics},
  volume       = {57},
  number       = {10},
  pages        = {105501},
  year         = {2024},
  doi          = {10.1088/1361-6455/ad3bff},
}

@article{Stellmer2014_SrRepump,
  author       = {Stellmer, Simon and Schreck, Florian},
  title        = {Metastable state repumping for compact ultracold strontium apparatus},
  journal      = {Physical Review A},
  volume       = {90},
  number       = {2},
  pages        = {022512},
  year         = {2014},
  month        = {August},
  doi          = {10.1103/PhysRevA.90.022512},
  publisher    = {American Physical Society}
}

@article{Coherent_Control_Over_High-Dimensiona_Ahmed,
  title = {Coherent Control Over the High-Dimensional Space of the Nuclear Spin of Alkaline-Earth Atoms},
  author = {Ahmed, H. and Litvinov, A. and Guesdon, P. and Mar\'echal, E. and Huckans, J.H. and Pasquiou, B. and Laburthe-Tolra, B. and Robert-de-Saint-Vincent, M.},
  journal = {PRX Quantum},
  volume = {6},
  issue = {2},
  pages = {020352},
  numpages = {15},
  year = {2025},
  month = {Jun},
  publisher = {American Physical Society},
  doi = {10.1103/PRXQuantum.6.020352},
}

@article{Su_FAST_2025,
  title = {Fast single atom imaging for optical lattice arrays},
  author = {Su, Lin and Douglas, Alexander and Szurek, Michal and Hébert, Anne H. and Krahn, Aaron and Groth, Robin and Phelps, Gregory A. and Markovi\'c, Ognjen and Greiner, Markus},
  journal = {Nature Communications},
  volume = {16},
  issue = {1},
  pages = {1017},
  numpages = {—},
  year = {2025},
  month = {Jan},
  publisher = {Springer Nature},
  doi = {10.1038/s41467-025-56305-y},
  url = {https://www.nature.com/articles/s41467-025-56305-y}
}

@article{Jackson_2020,
  author    = {Jackson, Niamh Christina and Hanley, Ryan Keith and Hill, Matthew and Leroux, Frédéric and Adams, Charles S. and Jones, Matthew Philip Austin},
  title     = {Number‑resolved imaging of ${}^{88}$Sr atoms in a long working distance optical tweezer},
  journal   = {SciPost Physics},
  volume    = {8},
  issue     = {3},
  pages     = {038},
  year      = {2020},
  month     = {Mar},
  publisher = {SciPost},
  doi       = {10.21468/SciPostPhys.8.3.038},
  url       = {https://scipost.org/SciPostPhys.8.3.038}
}

@article{Bluvstein_2024,
  author    = {Bluvstein, Dolev and Evered, Simon J. and Geim, Alexandra A. and Li, Sophie H. and Zhou, Hengyun and Manovitz, Tom and Ebadi, Sepehr and Cain, Madelyn and Kalinowski, Marcin and Hangleiter, Dominik and Bonilla Ataides, J. Pablo and Maskara, Nishad and Cong, Iris and Gao, Xun and Sales Rodriguez, Pedro and Karolyshyn, Thomas and Semeghini, Giulia and Gullans, Michael J. and Greiner, Markus and Vuletić, Vladan and Lukin, Mikhail D.},
  title     = {Logical quantum processor based on reconfigurable atom arrays},
  journal   = {Nature},
  volume    = {626},
  issue     = {7997},
  pages     = {58–65},
  year      = {2024},
  month     = {Feb},
  publisher = {Springer Nature},
  doi       = {10.1038/s41586-023-06927-3},
  url       = {https://doi.org/10.1038/s41586-023-06927-3}
}

@article{Norcia2018_Sr_Tweezers_GSC,
  title   = {Microscopic Control and Detection of Ultracold Strontium in Optical-Tweezer Arrays},
  author  = {Norcia, Matthew A. and Young, Aaron W. and Kaufman, Adam M.},
  journal = {Phys. Rev. X},
  volume  = {8},
  pages   = {041054},
  year    = {2018},
  doi     = {10.1103/PhysRevX.8.041054}
}

@article{boughdachi2025strontium,
  title = {Quasicontinuous sub-$\text{\ensuremath{\mu}}\mathrm{K}$ strontium source without a high-finesse cavity-stabilized laser},
  author = {Boughdachi, Sana and Heizenreder, Benedikt and Sitaram, Ananya and Dierikx, Erik and Xie, Yan and Klemann, Sander and Klop, Paul and Koelemeij, Jeroen and Wilk, Rafa\l{} and Schreck, Florian and Brodschelm, Andreas},
  journal = {Phys. Rev. Appl.},
  volume = {25},
  issue = {3},
  pages = {034053},
  numpages = {13},
  year = {2026},
  month = {Mar},
  publisher = {American Physical Society},
  doi = {10.1103/3gnx-4s97},
  url = {https://link.aps.org/doi/10.1103/3gnx-4s97}
}

@article{Atom_computing,
  title = {Iterative Assembly of ${}^{171}$$\mathrm{Yb}$ Atom Arrays with Cavity-Enhanced Optical Lattices},
  author = {Norcia, M. A. and Kim, H. and Cairncross, W. B. and Bloom, B. J. and others},
  journal = {PRX Quantum},
  volume = {5},
  issue = {3},
  pages = {030316},
  numpages = {13},
  year = {2024},
  month = {Jul},
  publisher = {American Physical Society},
  doi = {10.1103/PRXQuantum.5.030316},
}

@article{Li2025_fast_continuous_coherent,
  author  = {Yiyi Li and Yicheng Bao and Michael Peper and Chenyuan Li and Jeff D. Thompson},
  title   = {Fast, continuous and coherent atom replacement in a neutral atom qubit array},
  journal = {arXiv preprint},
  volume  = {arXiv:2506.15633},
  year    = {2025},
  url     = {https://doi.org/10.48550/arXiv.2506.15633}
}

@article{Chiu2025_continuous_3000_qubit,
  author  = {Chiu, Neng-Chun and Trapp, Elias C. and Guo, Jinen and Abobeih, Mohamed H. and Stewart, Luke M. and Hollerith, Simon and Stroganov, Pavel L. and Kalinowski, Marcin and Geim, Alexandra A. and Evered, Simon J. and Li, Sophie H. and Lyu, Xingjian and Peters, Lisa M. and Bluvstein, Dolev and Wang, Tout T. and Greiner, Markus and Vuletić, Vladan and Lukin, Mikhail D.},
  title   = {Continuous operation of a coherent 3,000-qubit system},
  journal = {Nature},
  volume  = {646},
  pages   = {1075--1080},
  year    = {2025},
  doi     = {10.1038/s41586-025-09596-6},
}

@article{Okamoto2025_OpticalPumping_448nm_Sr,
  title = {Analyzing the optical pumping on the $5s4d^{1}D_{2}\text{\ensuremath{-}}5s8p^{1}P_{1}$ transition in a magneto-optical trap of Sr atoms},
  author = {Okamoto, Naohiro and Aoki, Takatoshi and Torii, Yoshio},
  journal = {Phys. Rev. A},
  volume = {112},
  issue = {4},
  pages = {043118},
  numpages = {7},
  year = {2025},
  month = {Oct},
  publisher = {American Physical Society},
  doi = {10.1103/j7s7-b7mj},
  url = {https://link.aps.org/doi/10.1103/j7s7-b7mj}
}

@Article{Knottnerus2025_ParallelAssembly_Tweezers_SLM,
	title={{Parallel assembly of neutral atom arrays with an SLM using linear phase interpolation}},
	author={Ivo H. A. Knottnerus and Yu Chih Tseng and Alexander Urech and Robert J. C. Spreeuw and Florian Schreck},
	journal={SciPost Phys.},
	volume={19},
	pages={118},
	year={2025},
	publisher={SciPost},
	doi={10.21468/SciPostPhys.19.4.118},
	url={https://scipost.org/10.21468/SciPostPhys.19.4.118},
}

@article{Lin2025_AIAssembly_PRL,
  author  = {Rui Lin and Han-Sen Zhong and You Li and Zhang-Rui Zhao and Le-Tian Zheng and Tai-Ran Hu and Hong-Ming Wu and Zhan Wu and Wei-Jie Ma and Yan Gao and Yi-Kang Zhu and Zhao-Feng Su and Wan-Li Ouyang and Yu-Chen Zhang and Jun Rui and Ming-Cheng Chen and Chao-Yang Lu and Jian-Wei Pan},
  title   = {AI-Enabled Parallel Assembly of Thousands of Defect-Free Neutral Atom Arrays},
  journal = {Phys. Rev. Lett.},
  volume  = {135},
  pages   = {060602},
  year    = {2025},
  url     = {https://doi.org/10.1103/PhysRevLett.135.060602}
}

@article{Rodrigo_MOT_2021,
  title = {Steady-state magneto-optical trap of fermionic strontium on a narrow-line transition},
  author = {Escudero, Rodrigo Gonz\'alez and Chen, Chun-Chia and Bennetts, Shayne and Pasquiou, Benjamin and Schreck, Florian},
  journal = {Phys. Rev. Res.},
  volume = {3},
  issue = {3},
  pages = {033159},
  numpages = {11},
  year = {2021},
  month = {Aug},
  publisher = {American Physical Society},
  doi = {10.1103/PhysRevResearch.3.033159},
  url = {https://link.aps.org/doi/10.1103/PhysRevResearch.3.033159}
}

@book{Haynes2016,
  author    = {W. M. Haynes},
  title     = {CRC Handbook of Chemistry and Physics},
  edition   = {97th},
  publisher = {CRC Press},
  address   = {Boca Raton, FL},
  year      = {2016},
  pages     = {2670},
  url       = {https://www.crcpress.com/CRC-Handbook-of-Chemistry-and-Physics-97th-Edition/Haynes/p/book/9781498754293}
}

@article{Pucher2025_88Sr_Reference,
  author  = {Sebastian Pucher and Sofus Laguna Kristensen and Ronen M. Kroeze},
  title   = {${}^{88}$Sr Reference Data},
  journal = {arXiv preprint},
  volume  = {arXiv:2507.10487},
  year    = {2025},
  url     = {https://arxiv.org/pdf/2507.10487}
}

@article{Gross2024_HardwareEfficient,
  author  = {Jonathan A. Gross and Cl\'ement Godfrin and Alexandre Blais and Eva Dupont-Ferrier},
  title   = {Hardware-efficient error-correcting codes for large nuclear spins},
  journal = {Phys. Rev. Appl.},
  volume  = {22},
  issue   = {1},
  pages   = {014006},
  year    = {2024},
  url     = {https://link.aps.org/doi/10.1103/PhysRevApplied.22.014006}
}

@article{Campbell2014_EnhancedFaultTolerant_dLevel,
  author  = {Earl T. Campbell},
  title   = {Enhanced Fault-Tolerant Quantum Computing in $d$-Level Systems},
  journal = {Phys. Rev. Lett.},
  volume  = {113},
  issue   = {23},
  pages   = {230501},
  numpages= {5},
  year    = {2014},
  month   = {Dec},
  publisher = {American Physical Society},
  doi     = {10.1103/PhysRevLett.113.230501},
  url     = {https://link.aps.org/doi/10.1103/PhysRevLett.113.230501}
}

@article{Lim2023_Spin7_2Qudit,
  author  = {Sumin Lim and Junjie Liu and Arzhang Ardavan},
  title   = {Fault-tolerant qubit encoding using a spin-7/2 qudit},
  journal = {Phys. Rev. A},
  volume  = {108},
  issue   = {6},
  pages   = {062403},
  year    = {2023},
  url     = {https://link.aps.org/doi/10.1103/PhysRevA.108.062403}
}

@article{Lim2025_ErrorCorrectable_Qudit,
  author  = {Sumin Lim and Mikhail V. Vaganov and Junjie Liu and Arzhang Ardavan},
  title   = {Demonstrating experimentally the encoding and dynamics of an error-correctable logical qubit on a hyperfine-coupled nuclear spin qudit},
  journal = {Phys. Rev. Lett.},
  volume  = {134},
  issue   = {7},
  pages   = {070603},
  numpages= {7},
  year    = {2025},
  month   = {Feb},
  publisher = {American Physical Society},
  doi     = {10.1103/PhysRevLett.134.070603},
  url     = {https://link.aps.org/doi/10.1103/PhysRevLett.134.070603}
}

@article{Mezzadri2024_FaultTolerantQudit,
  author    = {Matteo Mezzadri and Alessandro Chiesa and Luca Lepori and Stefano Carretta},
  title     = {Fault-tolerant computing with single-qudit encoding in a molecular spin},
  journal   = {Mater. Horiz.},
  volume    = {11},
  pages     = {4961--4969},
  year      = {2024},
  doi       = {10.1039/D4MH00454J},
  url       = {https://pubs.rsc.org/en/content/articlepdf/2024/mh/d4mh00454j}
}

@article{Yan2025_ClockLaserQudit,
title = {High-Power Clock Laser Spectrally Tailored for High-Fidelity Quantum State Engineering},
  author = {Yan, Lingfeng and Lannig, Stefan and Milner, William R. and Frankel, Max N. and Lewis, Ben and Lee, Dahyeon and Kim, Kyungtae and Ye, Jun},
  journal = {Phys. Rev. X},
  volume = {15},
  issue = {3},
  pages = {031055},
  numpages = {19},
  year = {2025},
  month = {Aug},
  publisher = {American Physical Society},
  doi = {10.1103/qw53-8b8r},
  url = {https://link.aps.org/doi/10.1103/qw53-8b8r}
}

@article{Milner2025_SuperexchangeClock,
  author  = {William R. Milner and Lingfeng Yan and Stefan Lannig and William R. Milner and Max N. Frankel and Ben Lewis and Dahyeon Lee and Kyungtae Kim and Jun Ye},
  title   = {Coherent evolution of superexchange interaction in seconds-long optical clock spectroscopy},
  journal = {Science},
  volume  = {380},
  issue   = {6634},
  pages   = {1234--1238},
  year    = {2025},
  doi     = {10.1126/science.ado5987},
  url     = {https://doi.org/10.1126/science.ado5987}
}

@article{Kim2025_WannierStarkClock,
   title = {Atomic Coherence of 2 Minutes and Instability of $1.5\ifmmode\times\else\texttimes\fi{}{10}^{\ensuremath{-}18}$ at 1 s in a Wannier-Stark Lattice Clock},
  author = {Kim, Kyungtae and Aeppli, Alexander and Warfield, William and Chu, Anjun and Rey, Ana Maria and Ye, Jun},
  journal = {Phys. Rev. Lett.},
  volume = {135},
  issue = {10},
  pages = {103601},
  numpages = {7},
  year = {2025},
  month = {Sep},
  publisher = {American Physical Society},
  doi = {10.1103/3wtv-sty2},
  url = {https://link.aps.org/doi/10.1103/3wtv-sty2}
}

@article{yu_excess_2023,
  title   = {Excess Noise and Photoinduced Effects in Highly Reflective Crystalline Mirror Coatings},
  author  = {Yu, Jialiang and Häfner, Sebastian and Legero, Thomas and Herbers, Sofia and Nicolodi, Daniele and Ma, Chun Yu and Riehle, Fritz and Sterr, Uwe and Kedar, Dhruv and Robinson, John M. and Oelker, Eric and Ye, Jun},
  journal = {Phys. Rev. X},
  volume  = {13},
  number  = {4},
  pages   = {041002},
  year    = {2023},
  doi     = {10.1103/PhysRevX.13.041002}
}

@article{kedar_frequency_2023,
  title   = {Frequency Stability of Cryogenic Silicon Cavities with Semiconductor Crystalline Coatings},
  author  = {Kedar, Dhruv and Yu, Jialiang and Oelker, Eric and Staron, Alexander and Milner, William R. and Robinson, John M. and Legero, Thomas and Riehle, Fritz and Sterr, Uwe and Ye, Jun},
  journal = {Optica},
  volume  = {10},
  number  = {4},
  pages   = {464--471},
  year    = {2023},
  doi     = {10.1364/OPTICA.479462}
}

@article{muniz_high-fidelity_2024,
  author  = {J. A. Muniz and M. Stone and D. T. Stack and M. Jaffe and J. M. Kindem and L. Wadleigh and E. Zalys-Geller and X. Zhang and C.-A. Chen and M. A. Norcia and et al.},
  title   = {High-Fidelity Universal Gates in the ${}^{171}\mathrm{Yb}$ Ground-State Nuclear-Spin Qubit},
  journal = {PRX Quantum},
  volume  = {6},
  number  = {2},
  pages   = {020334},
  year    = {2025},
  doi     = {10.1103/PRXQuantum.6.020334},
  url     = {https://link.aps.org/doi/10.1103/PRXQuantum.6.020334}
}

@book{atomic,
    author = "J.T.M. Walraven",
    title = "Atomic physics lectures",
    publisher = "University of Amsterdam",
    year = "2025",
    url = "https://staff.fnwi.uva.nl/j.t.m.walraven/walraven/Publications_files/2021-AtomicPhysics.pdf"
}

@article{polkovnikov2010phase,
  author  = {A. Polkovnikov},
  title   = {Phase space representation of quantum dynamics},
  journal = {Annals of Physics},
  volume  = {325},
  pages   = {1790–1852},
  year    = {2010},
  doi     = {10.1016/j.aop.2010.02.006},
  url     = {https://www.sciencedirect.com/science/article/abs/pii/S0003491610000382}
}

@article{Huie2025_threequbit,
  title = {Three-Qubit Encoding in Ytterbium-171 Atoms for Simulating 1+1D Quantum Chromodynamics},
  author = {Huie, William and Conefrey-Shinozaki, Cianan and Jia, Zhubing and Draper, Patrick and Covey, Jacob P.},
  journal = {PRX Quantum},
  volume = {7},
  issue = {1},
  pages = {010327},
  numpages = {23},
  year = {2026},
  month = {Feb},
  publisher = {American Physical Society},
  doi = {10.1103/28h7-jl1v},
  url = {https://link.aps.org/doi/10.1103/28h7-jl1v}
}

@article{Verstraten2025_control,
  author       = {Robin C. Verstraten and Ivo H. A. Knottnerus and Yu Chih Tseng and Alexander Urech and Tiago Santiago do Espirito Santo and Vinicius Zampronio and Florian Schreck and Robert J. C. Spreeuw and Cristiane Morais Smith},
  title        = {Control of single spin-flips in a Rydberg atomic fractal},
  journal      = {arXiv preprint},
  volume       = {arXiv:2509.03514},
  year         = {2025},
  url          = {https://doi.org/10.48550/arXiv.2509.03514}
}

@article{Modeling_swadheen,
  title = {Modeling of a continuous superradiant laser on the sub-mHz ${}^{1}{\mathrm{S}}_{0}\ensuremath{\rightarrow}{}^{3}{\mathrm{P}}_{0}$ transition in neutral strontium-88},
  author = {Dubey, Swadheen and Kazakov, Georgy A. and Heizenreder, Benedikt and Zhou, Sheng and Bennetts, Shayne and Sch\"affer, Stefan Alaric and Sitaram, Ananya and Schreck, Florian},
  journal = {Phys. Rev. Res.},
  volume = {7},
  issue = {1},
  pages = {013292},
  numpages = {20},
  year = {2025},
  month = {Mar},
  publisher = {American Physical Society},
  doi = {10.1103/PhysRevResearch.7.013292},
}

@article{Khaneja_Optimal_control,
title = {Optimal control of coupled spin dynamics: design of NMR pulse sequences by gradient ascent algorithms},
journal = {Journal of Magnetic Resonance},
volume = {172},
number = {2},
pages = {296-305},
year = {2005},
issn = {1090-7807},
doi = {https://doi.org/10.1016/j.jmr.2004.11.004},
url = {https://www.sciencedirect.com/science/article/pii/S1090780704003696},
author = {Navin Khaneja and Timo Reiss and Cindie Kehlet and Thomas Schulte-Herbrüggen and Steffen J. Glaser},
}

@article{Burshtein2025_robust,
  title = {Robust control and entanglement of qudits in neutral atom arrays},
  author = {Burshtein, Amir and Fraenkel, Shachar and Goldstein, Moshe and Finkelstein, Ran},
  journal = {Phys. Rev. Res.},
  volume = {8},
  issue = {1},
  pages = {013055},
  numpages = {16},
  year = {2026},
  month = {Jan},
  publisher = {American Physical Society},
  doi = {10.1103/4xcd-wyxx},
  url = {https://link.aps.org/doi/10.1103/4xcd-wyxx}
}

@article{Naydenov_Dynamical_decoupling,
  title = {Dynamical decoupling of a single-electron spin at room temperature},
  author = {Naydenov, Boris and Dolde, Florian and Hall, Liam T. and Shin, Chang and Fedder, Helmut and Hollenberg, Lloyd C. L. and Jelezko, Fedor and Wrachtrup, J\"org},
  journal = {Phys. Rev. B},
  volume = {83},
  issue = {8},
  pages = {081201},
  numpages = {4},
  year = {2011},
  month = {Feb},
  publisher = {American Physical Society},
  doi = {10.1103/PhysRevB.83.081201},
  url = {https://link.aps.org/doi/10.1103/PhysRevB.83.081201}
}

@article{de_Lange_Multipulse_Sensing_Sequences,
  title = {Single-Spin Magnetometry with Multipulse Sensing Sequences},
  author = {de Lange, G. and Rist\`e, D. and Dobrovitski, V. V. and Hanson, R.},
  journal = {Phys. Rev. Lett.},
  volume = {106},
  issue = {8},
  pages = {080802},
  numpages = {4},
  year = {2011},
  month = {Feb},
  publisher = {American Physical Society},
  doi = {10.1103/PhysRevLett.106.080802},
  url = {https://link.aps.org/doi/10.1103/PhysRevLett.106.080802}
}

@article{spreeuw2020off,
  author = {Spreeuw, R. J. C.},
  title = {Off-axis dipole forces in optical tweezers by an optical analog of the Magnus effect},
  journal = {Phys. Rev. Lett.},
  volume = {125},
  number = {23},
  pages = {233201},
  year = {2020},
  doi = {10.1103/PhysRevLett.125.233201},
  url = {https://journals.aps.org/prl/abstract/10.1103/PhysRevLett.125.233201}
}

@phdthesis{urech2023single,
  author       = {A. A. Urech},
  title        = {Single strontium atoms in optical tweezers},
  school       = {University of Amsterdam},
  address      = {Amsterdam, The Netherlands},
  year         = {2023},
  url          = {https://www.strontiumbec.com/StrontiumLab/Theses/Alex_Urech_PhD_thesis.pdf}
}

@article{qutip1,
title = {QuTiP: An open-source Python framework for the dynamics of open quantum systems},
journal = {Computer Physics Communications},
volume = {183},
number = {8},
pages = {1760-1772},
year = {2012},
issn = {0010-4655},
doi = {https://doi.org/10.1016/j.cpc.2012.02.021},
url = {https://www.sciencedirect.com/science/article/pii/S0010465512000835},
author = {J.R. Johansson and P.D. Nation and Franco Nori}
}

@article{qutip2,
title = {QuTiP 2: A Python framework for the dynamics of open quantum systems},
journal = {Computer Physics Communications},
volume = {184},
number = {4},
pages = {1234-1240},
year = {2013},
issn = {0010-4655},
doi = {https://doi.org/10.1016/j.cpc.2012.11.019},
url = {https://www.sciencedirect.com/science/article/pii/S0010465512003955},
author = {J.R. Johansson and P.D. Nation and Franco Nori}
}

@article{qutip3,
title = {QuTiP 5: The Quantum Toolbox in Python},
journal = {Physics Reports},
volume = {1153},
pages = {1-62},
year = {2026},
issn = {0370-1573},
doi = {https://doi.org/10.1016/j.physrep.2025.10.001},
url = {https://www.sciencedirect.com/science/article/pii/S0370157325002704},
author = {Neill Lambert and Eric Giguère and Paul Menczel and Boxi Li and Patrick Hopf and Gerardo Suárez and Marc Gali and Jake Lishman and Rushiraj Gadhvi and Rochisha Agarwal and Asier Galicia and Nathan Shammah and Paul Nation and J.R. Johansson and Shahnawaz Ahmed and Simon Cross and Alexander Pitchford and Franco Nori}
}

@article{Okamoto2025DirectMeasurement,
  title = {Direct measurement of the $5s5p^{1}{P}_{1}\ensuremath{\rightarrow}5s4d^{1}{D}_{2}$ decay rate in strontium},
  author = {Okamoto, Naohiro and Aoki, Takatoshi and Torii, Yoshio},
  journal = {Phys. Rev. A},
  volume = {113},
  issue = {6},
  pages = {L060803},
  numpages = {5},
  year = {2026},
  month = {Jun},
  publisher = {American Physical Society},
  doi = {10.1103/8b7c-l2jl},
  url = {https://link.aps.org/doi/10.1103/8b7c-l2jl}
}

@article{Wu2022_ErasureConversion_AlkalineEarth,
  title        = {Erasure conversion for fault-tolerant quantum computing in alkaline earth Rydberg atom arrays},
  author       = {Wu, Yue and Kolkowitz, Shimon and Puri, Shruti and Thompson, Jeff D.},
  journal      = {Nature Communications},
  volume       = {13},
  pages        = {4657},
  year         = {2022},
  doi          = {10.1038/s41467-022-32094-6},
  url          = {https://www.nature.com/articles/s41467-022-32094-6}
}

@article{Scholl2023_ErasureConversion_Rydberg,
  title        = {Erasure conversion in a high-fidelity Rydberg quantum simulator},
  author       = {Scholl, Pascal and Shaw, Adam L. and Tsai, Richard Bing-Shiun and Finkelstein, Ran and Choi, Joonhee and Endres, Manuel},
  journal      = {Nature},
  volume       = {622},
  issue        = {7982},
  pages        = {273–278},
  year         = {2023},
  doi          = {10.1038/s41586-023-06516-4},
  url          = {https://www.nature.com/articles/s41586-023-06516-4}
}

@misc{3P2_2025data,
  author       = {B. Heizenreder and B. Gerritsen and K. Fouka and F. Schreck and A. Safavi Naini and A. Urech},
  title        = {Engineering Zeeman-manifold quintets using state-dependent light shifts in neutral atoms -- Data repository},
  year         = {2025},
  doi          = {10.21942/uva.30895847},
  note         = {Data repository}
}

@article{Zhang2024RecoilFreeGates,
  author  = {Zhao Zhang and Léo Van Damme and Marco Rossignolo and Lorenzo Festa and Max Melchner and Robin Eberhard and Dimitrios Tsevas and Kevin Mours and Eran Reches and Johannes Zeiher and et al.},
  title   = {Recoil-free Quantum Gates with Optical Qubits},
  journal = {arXiv preprint},
  volume  = {arXiv:2408.04622},
  year    = {2024},
  url     = {https://doi.org/10.48550/arXiv.2408.04622}
}

@article{Panelli_Doppler-free_three-photon_2025,
  title = {Doppler-free three-photon spectroscopy on narrow-line optical transitions},
  author = {Panelli, Guglielmo and Burd, Shaun C. and Porter, Erik J. and Kasevich, Mark},
  journal = {Phys. Rev. A},
  volume = {111},
  issue = {3},
  pages = {033112},
  numpages = {9},
  year = {2025},
  month = {Mar},
  publisher = {American Physical Society},
  doi = {10.1103/PhysRevA.111.033112},
  url = {https://link.aps.org/doi/10.1103/PhysRevA.111.033112}
}

@article{Barker_Three_photon_2016,
  title = {Three-photon process for producing a degenerate gas of metastable alkaline-earth-metal atoms},
  author = {Barker, D. S. and Pisenti, N. C. and Reschovsky, B. J. and Campbell, G. K.},
  journal = {Phys. Rev. A},
  volume = {93},
  issue = {5},
  pages = {053417},
  numpages = {8},
  year = {2016},
  month = {May},
  publisher = {American Physical Society},
  doi = {10.1103/PhysRevA.93.053417},
  url = {https://link.aps.org/doi/10.1103/PhysRevA.93.053417}
}

@article{albano2002squeezed,
  author  = {L. Albano and D. F. Mundarain and J. Stephany},
  title   = {On the squeezed number states and their phase-space representations},
  journal = {Journal of Optics B: Quantum and Semiclassical Optics},
  volume  = {4},
  pages   = {352--357},
  year    = {2002},
  doi     = {10.1088/1464-4266/4/5/319},
  url     = {https://iopscience.iop.org/article/10.1088/1464-4266/4/5/319}
}

@Article{Meth2025,
author={Meth, Michael
and Zhang, Jinglei
and Haase, Jan F.
and Edmunds, Claire
and Postler, Lukas
and Jena, Andrew J.
and Steiner, Alex
and Dellantonio, Luca
and Blatt, Rainer
and Zoller, Peter
and Monz, Thomas
and Schindler, Philipp
and Muschik, Christine
and Ringbauer, Martin},
title={Simulating two-dimensional lattice gauge theories on a qudit quantum computer},
journal={Nature Physics},
year={2025},
month={Apr},
day={01},
volume={21},
number={4},
pages={570-576},
issn={1745-2481},
doi={10.1038/s41567-025-02797-w},
url={https://doi.org/10.1038/s41567-025-02797-w}
}

@article{Gonzalez2022,
  title = {Hardware Efficient Quantum Simulation of Non-Abelian Gauge Theories with Qudits on Rydberg Platforms},
  author = {Gonz\'alez-Cuadra, Daniel and Zache, Torsten V. and Carrasco, Jose and Kraus, Barbara and Zoller, Peter},
  journal = {Phys. Rev. Lett.},
  volume = {129},
  issue = {16},
  pages = {160501},
  numpages = {8},
  year = {2022},
  month = {Oct},
  publisher = {American Physical Society},
  doi = {10.1103/PhysRevLett.129.160501},
  url = {https://link.aps.org/doi/10.1103/PhysRevLett.129.160501}
}

@article{Gaz2025,
  title = {Quantum simulation of non-Abelian lattice gauge theories: A variational approach to D8 with dynamical matter},
  author = {Gaz, Emanuele and Popov, Pavel P. and Pardo, Guy and Lewenstein, Maciej and Hauke, Philipp and Zohar, Erez},
  journal = {Phys. Rev. Res.},
  volume = {7},
  issue = {3},
  pages = {033012},
  numpages = {20},
  year = {2025},
  month = {Jul},
  publisher = {American Physical Society},
  doi = {10.1103/l8b6-h5s5},
  url = {https://link.aps.org/doi/10.1103/l8b6-h5s5}
}

@article{Illa2023,
  title = {Quantum simulations of SO(5) many-fermion systems using qudits},
  author = {Illa, Marc and Robin, Caroline E. P. and Savage, Martin J.},
  journal = {Phys. Rev. C},
  volume = {108},
  issue = {6},
  pages = {064306},
  numpages = {31},
  year = {2023},
  month = {Dec},
  publisher = {American Physical Society},
  doi = {10.1103/PhysRevC.108.064306},
  url = {https://link.aps.org/doi/10.1103/PhysRevC.108.064306}
}

@article{Ciavarella2021,
  title = {Trailhead for quantum simulation of SU(3) Yang-Mills lattice gauge theory in the local multiplet basis},
  author = {Ciavarella, Anthony and Klco, Natalie and Savage, Martin J.},
  journal = {Phys. Rev. D},
  volume = {103},
  issue = {9},
  pages = {094501},
  numpages = {45},
  year = {2021},
  month = {May},
  publisher = {American Physical Society},
  doi = {10.1103/PhysRevD.103.094501},
  url = {https://link.aps.org/doi/10.1103/PhysRevD.103.094501}
}

@article{Popov2024,
  title = {Variational quantum simulation of U(1) lattice gauge theories with qudit systems},
  author = {Popov, Pavel P. and Meth, Michael and Lewestein, Maciej and Hauke, Philipp and Ringbauer, Martin and Zohar, Erez and Kasper, Valentin},
  journal = {Phys. Rev. Res.},
  volume = {6},
  issue = {1},
  pages = {013202},
  numpages = {19},
  year = {2024},
  month = {Feb},
  publisher = {American Physical Society},
  doi = {10.1103/PhysRevResearch.6.013202},
  url = {https://link.aps.org/doi/10.1103/PhysRevResearch.6.013202}
}

@article{Lanyon2009,
author={Lanyon, Benjamin P.
and Barbieri, Marco
and Almeida, Marcelo P.
and Jennewein, Thomas
and Ralph, Timothy C.
and Resch, Kevin J.
and Pryde, Geoff J.
and O'Brien, Jeremy L.
and Gilchrist, Alexei
and White, Andrew G.},
title={Simplifying quantum logic using higher-dimensional Hilbert spaces},
journal={Nature Physics},
year={2009},
month={Feb},
day={01},
volume={5},
number={2},
pages={134-140},
issn={1745-2481},
doi={10.1038/nphys1150},
url={https://doi.org/10.1038/nphys1150}
}

@article{Thompson_gradients,
  title = {Coherence and Raman Sideband Cooling of a Single Atom in an Optical Tweezer},
  author = {Thompson, J. D. and Tiecke, T. G. and Zibrov, A. S. and Vuleti\ifmmode \acute{c}\else \'{c}\fi{}, V. and Lukin, M. D.},
  journal = {Phys. Rev. Lett.},
  volume = {110},
  issue = {13},
  pages = {133001},
  numpages = {5},
  year = {2013},
  month = {Mar},
  publisher = {American Physical Society},
  doi = {10.1103/PhysRevLett.110.133001},
  url = {https://link.aps.org/doi/10.1103/PhysRevLett.110.133001}
}

@article{chew2024ultraprecise,
  title={Ultraprecise holographic optical tweezer array},
  author={Chew, Y Torii and Poitrinal, Martin and Tomita, Takafumi and Kitade, Sota and Mauricio, Jorge and Ohmori, Kenji and de L{\'e}s{\'e}leuc, Sylvain},
  journal={Physical Review A},
  volume={110},
  number={5},
  pages={053518},
  year={2024},
  publisher={APS}
}

@article{Ma_2026,
  title = {Beyond qubits: Multilevel quantum sensing for dark matter},
  author = {Ma, Xiaolin and Takhistov, Volodymyr and Mizuochi, Norikazu and Herbschleb, Ernst David},
  journal = {Phys. Rev. A},
  pages = {},
  year = {2026},
  month = {May},
  publisher = {American Physical Society},
  doi = {10.1103/zvzb-yv67},
  url = {https://link.aps.org/doi/10.1103/zvzb-yv67}
}

@article{Javaherian_2025,
  title = {Quantum noise spectroscopy by qudit spectators},
  author = {Javaherian, Clara and Ferrie, Chris},
  journal = {Phys. Rev. A},
  volume = {112},
  issue = {3},
  pages = {032411},
  numpages = {41},
  year = {2025},
  month = {Sep},
  publisher = {American Physical Society},
  doi = {10.1103/drc2-8qbg},
  url = {https://link.aps.org/doi/10.1103/drc2-8qbg}
}

@article{Calajo_2024,
  title = {Digital Quantum Simulation of a (1+1)D SU(2) Lattice Gauge Theory with Ion Qudits},
  author = {Calaj\'o, Giuseppe and Magnifico, Giuseppe and Edmunds, Claire and Ringbauer, Martin and Montangero, Simone and Silvi, Pietro},
  journal = {PRX Quantum},
  volume = {5},
  issue = {4},
  pages = {040309},
  numpages = {20},
  year = {2024},
  month = {Oct},
  publisher = {American Physical Society},
  doi = {10.1103/PRXQuantum.5.040309},
  url = {https://link.aps.org/doi/10.1103/PRXQuantum.5.040309}
}

@article{Kaubruegger2023,
  title = {Optimal and Variational Multiparameter Quantum Metrology and Vector-Field Sensing},
  author = {Kaubruegger, Raphael and Shankar, Athreya and Vasilyev, Denis V. and Zoller, Peter},
  journal = {PRX Quantum},
  volume = {4},
  issue = {2},
  pages = {020333},
  numpages = {21},
  year = {2023},
  month = {Jun},
  publisher = {American Physical Society},
  doi = {10.1103/PRXQuantum.4.020333},
  url = {https://link.aps.org/doi/10.1103/PRXQuantum.4.020333}
}

@article{Ahmed2025,
  title = {Coherent Control Over the High-Dimensional Space of the Nuclear Spin of Alkaline-Earth Atoms},
  author = {Ahmed, H. and Litvinov, A. and Guesdon, P. and Mar\'echal, E. and Huckans, J.H. and Pasquiou, B. and Laburthe-Tolra, B. and Robert-de-Saint-Vincent, M.},
  journal = {PRX Quantum},
  volume = {6},
  issue = {2},
  pages = {020352},
  numpages = {15},
  year = {2025},
  month = {Jun},
  publisher = {American Physical Society},
  doi = {10.1103/PRXQuantum.6.020352},
  url = {https://link.aps.org/doi/10.1103/PRXQuantum.6.020352}
}

@article{Edmunds2025,
  title = {Symmetry-Protected Topological Haldane Phase on a Qudit Quantum Processor},
  author = {Edmunds, C.L. and Rico, E. and Arrazola, I. and Brennen, G.K. and Meth, M. and Blatt, R. and Ringbauer, M.},
  journal = {PRX Quantum},
  volume = {6},
  issue = {2},
  pages = {020349},
  numpages = {13},
  year = {2025},
  month = {Jun},
  publisher = {American Physical Society},
  doi = {10.1103/PRXQuantum.6.020349},
  url = {https://link.aps.org/doi/10.1103/PRXQuantum.6.020349}
}

@article{Mogerle2025,
  title = {Spin-1 Haldane Phase in a Chain of Rydberg Atoms},
  author = {M\"ogerle, J. and Brechtelsbauer, K. and Gea-Caballero, A.T. and Prior, J. and Emperauger, G. and Bornet, G. and Chen, C. and Lahaye, T. and Browaeys, A. and B\"uchler, H.P.},
  journal = {PRX Quantum},
  volume = {6},
  issue = {2},
  pages = {020332},
  numpages = {13},
  year = {2025},
  month = {May},
  publisher = {American Physical Society},
  doi = {10.1103/PRXQuantum.6.020332},
  url = {https://link.aps.org/doi/10.1103/PRXQuantum.6.020332}
}

@article{Wang2023,
  title = {Dissipative preparation and stabilization of many-body quantum states in a superconducting qutrit array},
  author = {Wang, Yunzhao and Snizhko, Kyrylo and Romito, Alessandro and Gefen, Yuval and Murch, Kater},
  journal = {Phys. Rev. A},
  volume = {108},
  issue = {1},
  pages = {013712},
  numpages = {16},
  year = {2023},
  month = {Jul},
  publisher = {American Physical Society},
  doi = {10.1103/PhysRevA.108.013712},
  url = {https://link.aps.org/doi/10.1103/PhysRevA.108.013712}
}

@article{Senko2015,
  title = {Realization of a Quantum Integer-Spin Chain with Controllable Interactions},
  author = {Senko, C. and Richerme, P. and Smith, J. and Lee, A. and Cohen, I. and Retzker, A. and Monroe, C.},
  journal = {Phys. Rev. X},
  volume = {5},
  issue = {2},
  pages = {021026},
  numpages = {9},
  year = {2015},
  month = {Jun},
  publisher = {American Physical Society},
  doi = {10.1103/PhysRevX.5.021026},
  url = {https://link.aps.org/doi/10.1103/PhysRevX.5.021026}
}

@article{Cohen2015,
  title = {Simulating the Haldane phase in trapped-ion spins using optical fields},
  author = {Cohen, I. and Richerme, P. and Gong, Z.-X. and Monroe, C. and Retzker, A.},
  journal = {Phys. Rev. A},
  volume = {92},
  issue = {1},
  pages = {012334},
  numpages = {10},
  year = {2015},
  month = {Jul},
  publisher = {American Physical Society},
  doi = {10.1103/PhysRevA.92.012334},
  url = {https://link.aps.org/doi/10.1103/PhysRevA.92.012334}
}

@article{Cohen2014,
  title = {Proposal for Verification of the Haldane Phase Using Trapped Ions},
  author = {Cohen, I. and Retzker, A.},
  journal = {Phys. Rev. Lett.},
  volume = {112},
  issue = {4},
  pages = {040503},
  numpages = {5},
  year = {2014},
  month = {Jan},
  publisher = {American Physical Society},
  doi = {10.1103/PhysRevLett.112.040503},
  url = {https://link.aps.org/doi/10.1103/PhysRevLett.112.040503}
}

@misc{lin2026,
      title={Single-Ensemble Multiparameter Squeezing with Qudits}, 
      author={Xiaoshui Lin and Chunlei Qu and Chong Zu and Chuanwei Zhang},
      year={2026},
      eprint={2605.26377},
      archivePrefix={arXiv},
      primaryClass={quant-ph},
      url={https://arxiv.org/abs/2605.26377}, 
}

@article{toptica_1064nm,
title = {Ultra-Low Noise, 80 W Laser System at 532 nm: Enabling Titanium-Sapphire Pumping and Advanced Semiconductor Inspection with Superior Noise Performance},
author = {Marcel Holtz and Simon Reinisch and Harald Rossmeier and Pierre Laygue and Konstantinos Simeonidis and Matthias Scholz},
journal = {Conference on Lasers and Electro-Optics/Europe (CLEO/Europe 2025) and European Quantum Electronics Conference (EQEC 2025)},
pages = {26},
publisher = {Optica Publishing Group},
year = {2025},
url = {https://opg.optica.org/abstract.cfm?URI=CLEO_Europe-2025-cd_p_26},
}

@Article{Gonzalez-Cuadra2025,
author={Gonz{\'a}lez-Cuadra, Daniel
and Hamdan, Majd
and Zache, Torsten V.
and Braverman, Boris
and Kornja{\v{c}}a, Milan
and Lukin, Alexander
and Cant{\'u}, Sergio H.
and Liu, Fangli
and Wang, Sheng-Tao
and Keesling, Alexander
and Lukin, Mikhail D.
and Zoller, Peter
and Bylinskii, Alexei},
title={Observation of string breaking on a (2 + 1)D Rydberg quantum simulator},
journal={Nature},
year={2025},
month={Jun},
day={01},
volume={642},
number={8067},
pages={321-326},
issn={1476-4687},
doi={10.1038/s41586-025-09051-6},
url={https://doi.org/10.1038/s41586-025-09051-6}
}
\end{document}